\documentclass[reprint,noeprint,superscriptaddress,amsmath,amssymb,prx,preprintnumbers]{revtex4-2}

\usepackage{mathtools}
\usepackage{times}
\usepackage[colorlinks=true,linkcolor=blue,urlcolor=blue,citecolor=blue]{hyperref}

\newcommand{\vect}[1]{\boldsymbol{#1}}

\usepackage{graphicx}
\usepackage{dcolumn}
\usepackage{bm,dsfont}
\usepackage{float}
\usepackage{braket} 
\usepackage{xcolor}
\usepackage{tikz}

\newcommand{\nn}{\nonumber}

\newcommand{\abs}[1]{\lvert#1\rvert}

\newcommand{\tr}{\mathrm{Tr}}

\newcommand{\HarvardPhysics}{Department of Physics, Harvard University, Cambridge, MA 02138, USA}

\newcommand{\MIT}{Center for Theoretical Physics, Massachusetts Institute of Technology, Cambridge, MA 02139, USA}

\newcommand{\Waterloo}{Department of Physics and Astronomy, University of Waterloo, Waterloo, ON, Canada}

\newcommand{\Perimeter}{Perimeter Institute for Theoretical Physics, Waterloo, ON, Canada}

\begin{document}
	\title{Bulk and Boundary Quantum Phase Transitions in a Square Rydberg Atom Array}
	\author{Marcin Kalinowski}
	\email{mkalinowski@g.harvard.edu}
	\affiliation{\HarvardPhysics}
	\author{Rhine Samajdar}
	\affiliation{\HarvardPhysics}
	\author{Roger G. Melko}
	\affiliation{\Waterloo}
	\affiliation{\Perimeter}
	\author{Mikhail D. Lukin}
	\affiliation{\HarvardPhysics}
	\author{Subir Sachdev}
	\affiliation{\HarvardPhysics}
	\affiliation{School of Natural Sciences, Institute for Advanced Study, Princeton NJ 08540, USA}
	\author{Soonwon Choi}
	\affiliation{\MIT}
	
	\date{\today}
	\preprint{MIT-CTP/5342}

	\begin{abstract}
    Motivated by recent experimental realizations of exotic phases of matter on programmable quantum simulators, 
	we carry out a comprehensive theoretical study of quantum phase transitions in a Rydberg atom array on a square lattice, with both open and periodic boundary conditions.
	In the bulk, we identify several first-order and continuous phase transitions by performing large-scale quantum Monte Carlo simulations and develop an analytical understanding of the nature of these transitions using the framework of Landau-Ginzburg-Wilson theory.
	Remarkably, we find that under open boundary conditions, the boundary itself  undergoes a second-order quantum phase transition, independent of the bulk. These results explain recent experimental observations and provide important new insights into both the adiabatic state preparation of novel quantum phases and quantum optimization using Rydberg atom array platforms. 
	\end{abstract}
	
	\maketitle
	
\setlength{\parskip}{1mm plus1mm minus 1mm}

\section{Introduction}
Rydberg atom arrays have recently emerged as a powerful  platform for programmable quantum simulation~\cite{bernien2017probing,Keesling.2019,de2019observation,Ebadi.2021,scholl2021quantum} and quantum information processing~\cite{Saffman.2010,levine2018high,Levine.2019,morgado2021quantum}.
Recent theoretical~\cite{fendley2004competing, samajdar2018numerical, whitsitt2018quantum, PhysRevLett.122.017205,chepiga2021kibble} and experimental work on this system has allowed for unprecedented new insights into a variety of quantum phases characterized by complex density-wave~\cite{Ebadi.2021, Samajdar_2020} or topological~\cite{Samajdar.2021,Verresen.2020,Semeghini.2021} orders. Moreover, Rydberg simulators allow for detailed studies of dynamics across quantum phase transitions (QPTs) and other quantum critical phenomena. Finally, they provide a natural many-body platform for exploring quantum advantage in solving combinatorial optimization problems~\cite{pichler2018quantum,pichler2018computational}. These advances motivate detailed quantitative understanding of 
the QPTs between complex phases in such systems, with realistic interactions and geometries. In particular, on even the simplest square 2D arrays, where a compendium of ordered phases was theoretically predicted~\cite{Samajdar_2020}---and probed both by experiments~\cite{Ebadi.2021} and approximate numerics~\cite{Felser.2021}---a thorough classification of the associated QPTs is still lacking.

\begin{figure}[tb]
	    \centering
	    \includegraphics[width=\linewidth]{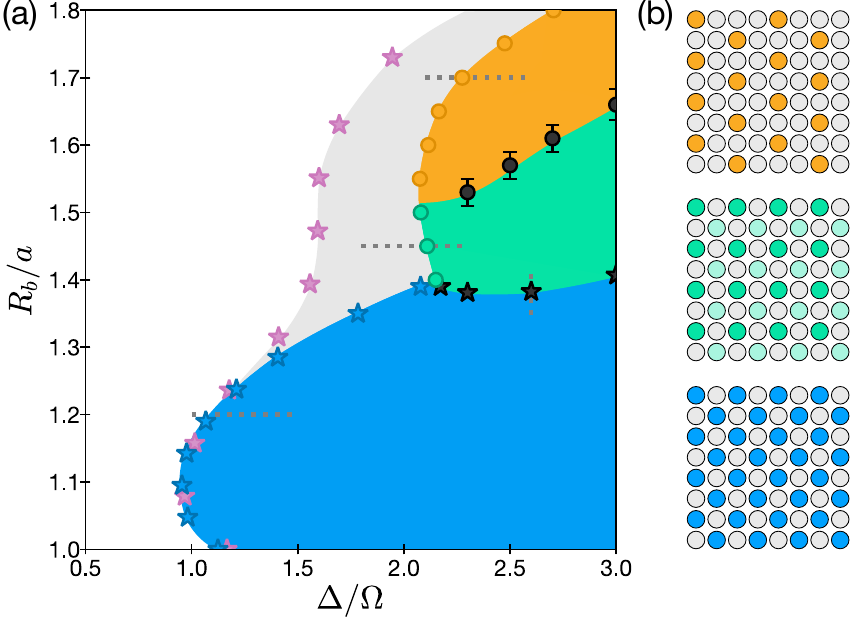}
	    \caption{
	    (a) Quantum phase diagram: first- and second-order QPTs are denoted by the circle and star markers, respectively. Purple stars mark boundary transitions in systems with OBC. The shaded area marks the region where the system lies in the disordered phase when using PBC but exhibits a boundary-ordered phase with OBC. The gray dashed lines indicate parameter ranges investigated in Figs.~\ref{fig:fig2} and~\ref{fig:fig3}.
	    (b) Schematic pictures of three distinct density-wave orders corresponding to the star (orange), striated (green), and checkerboard (blue) phases. Filled sites denote Rydberg excitations while gray sites represent ground-state atoms. 
	    }
	    \label{fig:fig1}
	\end{figure}

The many-body physics of  Rydberg atom arrays can be understood as resulting from two competing processes. 
On one hand, atoms in highly excited (Rydberg) states interact via strong van der Waals interactions~\cite{jaksch2000fast}, preventing neighboring atoms from simultaneously occupying the excited state---a mechanism known as the ``Rydberg blockade''~\cite{Lukin.2000}. On the other hand, a detuned laser field favors occupation of the Rydberg state, enticing the system to maximize the number of excited atoms~\cite{pohl2010dynamical}. This competition leads to a rich phase diagram~[Fig.~\ref{fig:fig1}(a)]. On a square lattice, when the blockade radius is comparable to the lattice spacing, the double occupancy of neighboring sites is highly suppressed, leading to a ``checkerboard'' pattern of excited atoms [Fig.~\ref{fig:fig1}(b)]. Upon increasing the blockade radius further, more complicated symmetry-breaking phases (namely, the ``striated'' and the ``star'') emerge~\cite{Samajdar_2020}. The transitions between these ordered phases (and to the disordered phase) can, in principle, be either continuous or first-order.
Crucially, the nature of these phase transitions dictates the efficacy of experimentally preparing the corresponding states via quasi-adiabatic dynamics in large systems.
Therefore, in order to utilize Rydberg atom arrays to probe different phases of matter or prepare the ground states of Hamiltonians encoding combinatorial optimization problems~\cite{pichler2018quantum}, it is essential to establish a quantitative understanding of the quantum critical points.

In this work, we employ both numerical and analytical methods to investigate the nature of these  QPTs in large systems with realistic, long-range interactions.
First, using large-scale Quantum Monte Carlo (QMC) simulations~\cite{Nakamura_2008}, we construct the phase diagram [Fig.~\ref{fig:fig1}(a)] and identify the nature of five distinct QPTs between the phases. This is possible  only because QMC can reach system sizes large enough for a careful finite-size scaling analysis under periodic boundary conditions (PBC), which are computationally difficult to realize in tensor-network-based  methods~\cite{Samajdar_2020,Felser.2021}. Interestingly, we find that while the QPTs from the disordered to the checkerboard phase and from the checkerboard to the striated phase are continuous, the transitions from the disordered to both the striated and the star phases are first-order.

Second, to understand the origin of the first-order transitions, we develop low-energy Landau-Ginzburg-Wilson (LGW) theories describing the QPTs in the system. Our analysis reveals the emergence of \emph{fluctuation-induced} first-order transitions, arising from the inaccessibility of stable fixed points in the renormalization group (RG) flow.
Additionally, we discover that systems  with open boundary conditions (OBC) can undergo boundary phase transitions [pink stars in Fig.~\ref{fig:fig1}(a)] independently from the bulk. 
Intuitively, the boundary transition is a consequence of the reduced connectivity near the boundary: fewer neighbors result in fewer blockade constraints and reduced frustration, allowing for easier ordering.

The presence of first-order phase transitions has significant ramifications for adiabatic state preparation in experiments: since the spectral gap often becomes exponentially small in the system size at a first-order transition point, it makes all adiabatic processes forbiddingly difficult.
This is in stark contrast to the successful experimental preparation of quantum phases reported in Ref.~\onlinecite{Ebadi.2021}.
We find that this  discrepancy is resolved by the boundary phase transition: for intermediate system sizes, the ordered boundary ``seeds” the bulk order and weakens the first-order transition, effectively enabling experimental adiabatic state preparation.
We also show that the observed phase diagram (specifically, the extent of the striated phase) is significantly affected by the presence of a boundary. 

The structure of this paper is as follows. First, we introduce our model of the Rydberg system in Sec.~\ref{sec:model}. Next, in Sec.~\ref{sec:transitions}, we numerically study the quantum phase transitions between the various phases in a system with PBC, and in Sec.~\ref{sec:lgw}, we present their unified field-theoretical description. Thereafter, in Sec.~\ref{sec:boundary}, we analyze the boundary ordering in a system with OBC. Finally, we summarize our conclusions in Sec.~\ref{sec:outlook}.
	
	\section{Model}\label{sec:model}
	We consider an array of Rydberg atoms on a square lattice interacting via the Hamiltonian
	\begin{equation}
	    H = \Omega\sum_{i<j} \left(R^{}_b/ R^{}_{ij}\right)^6 n^{}_i n^{}_j - \Delta\sum_i  n^{}_i +\frac{\Omega}{2}\sum_i \sigma^x_i,\label{eq:Ham}
	\end{equation}
	where $n^{}_i$\,$\equiv$\,$(\sigma^z_i+1)/2$ measures the Rydberg excitation density at site $i$, $R^{}_b$ is the so-called ``blockade radius''~\cite{Browaeys.2020} encapsulating the strength of the interactions, $R^{}_{ij}$ is the distance between sites $i$ and $j$, and $\sigma^x_i$ is the  Pauli $X$ operator at site $i$ representing a transverse field. $\Delta$ and $\Omega$\,$>$\,$0$ denote the detuning from the Rydberg state and the Rabi frequency, respectively. In practice, we adopt a finite cutoff for the interaction potential such that the interactions are set to 0 for
	$R_{ij}>R_0$.
	We systematically investigate different values of $R_0$ and choose an optimal value, $R_0 = 4a$ (where $a$ is the lattice spacing), that maximizes the computational efficiency without significantly affecting numerical results for given system sizes (the role of the cutoff distance is elaborated on in Appendix~\ref{app:cutoff}).

To characterize the phases and the transitions between them, we focus on the corresponding order parameters, defined by the symmetrized Fourier transform of the Rydberg density as $F(k_x,k_y)$\,$\equiv$\,$[\tilde{F}(k_x,k_y)$\,$+$\,$\tilde{F}(k_y,k_x)]$\,$/$\,$2$ at momentum $(k_x,k_y)$~\cite{Samajdar_2020,Ebadi.2021} with
	\begin{align}
	    \tilde{F}(k_x,k_y) &= \frac{1}{N_a}\sum_j n_j \,\exp \left[i(k_x,k_y)\cdot(x_j,y_j)\right], 
	\end{align}
	where $N_a$ is the total number of atoms. The checkerboard, striated, and star order parameters correspond to $F(\pi,\pi)$, $F(0,\pi)$, and $F(\pi,\pi/2)$, respectively. Additionally, we compute the Binder ratio of each order parameter $F$ as $U_4(F)$\,$\equiv$\,$(3$\,$-$\,$\braket{F^4}/\braket{F^2}^2)/2$~\cite{Binder.1981}, which is system-size independent at the quantum critical point of a second-order transition.
		Another useful observable is the average Rydberg density $n$\,$=$\,$(1/N_a)\sum_i n_i$---this is a first derivative of the free energy, so any sharp behavior of this quantity across a phase boundary may signal a first-order transition.

	\section{Numerical study of phase transitions}\label{sec:transitions}
	We begin by numerically examining the QPTs between the four phases in the considered phase diagram: disordered, checkerboard, striated, and star [Fig.~\ref{fig:fig1}(a)]. The disordered phase does not break any symmetries, while the checkerboard and striated phases break $\mathbb{Z}_2$ and $\mathbb{Z}_2$\,$\times$\,$ \mathbb{Z}_2$ translational symmetries, respectively. The star phase breaks both the $ \mathbb{Z}_2$ symmetry, and the $C_4$ rotational symmetry.

	For our numerical simulations, we adapt a QMC algorithm, based on the continuous imaginary-time representation~\cite{QMCFootnote}; the algorithm is local in space but nonlocal in the imaginary-time direction~\cite{Nakamura_2008}. We found that, for our system, this QMC method performs better than the conventional stochastic series expansion algorithm with cluster updates~\cite{Sandvik.2003}.
Our full QMC approach is detailed in Appendix~\ref{app:qmc}.

	\begin{figure}
	    \centering
	    \includegraphics[width=\linewidth]{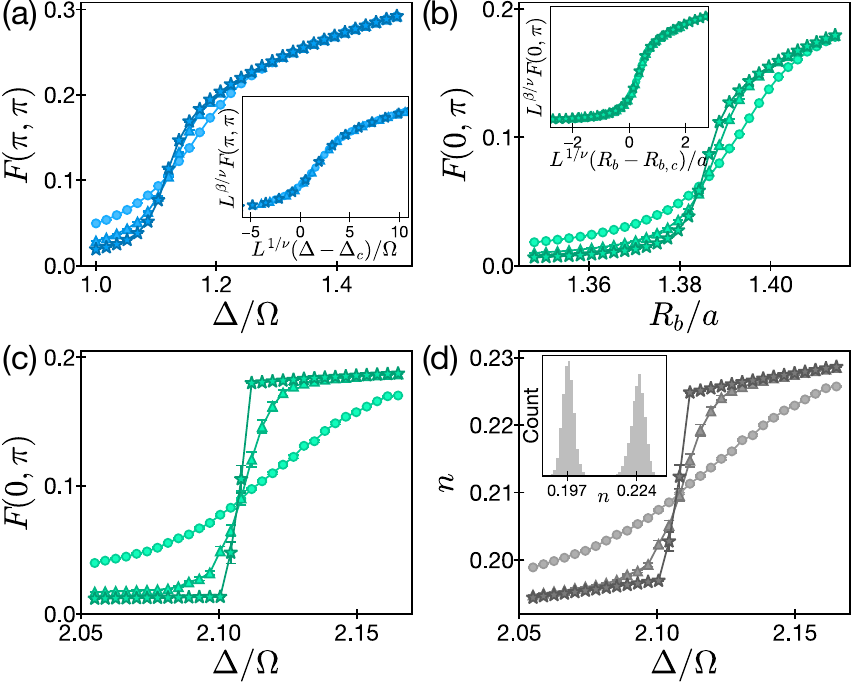}
	    \caption{Phase transitions between the disordered, checkerboard, and striated phases. Markers correspond to sizes $L$\,$=$\,$8$ (circle), 12 (triangle), 16 (star) with increasing color intensity. The order parameter of the checkerboard (striated) phase across the disordered--checkerboard (checkerboard--striated) boundary shows a second-order QPT in panel a (b), with its universal collapse presented in the inset. (c, d) The transition between the disordered and striated phases shows distinct signatures of a first-order transition: sharp jumps in the order parameter (c), and the Rydberg density (d) with a double-peaked distribution of QMC measurements at the transition.}
	    \label{fig:fig2}
	\end{figure}
	
	First, we study the transition from the disordered to the checkerboard phase. To this end, we calculate [Fig.~\ref{fig:fig2}(a)] the order parameter $F(\pi,\pi)$ and its Binder ratio across a range of detunings at a fixed value of the blockade radius $R_b$\,$=$\,$1.2$ [gray line in Fig.~\ref{fig:fig1}(a)]. We observe a smooth behavior of the Rydberg density $n$ and notice the Binder ratio crossing at a single point for multiple system sizes $N_a$\,$=$\,$L$\,$\times$\,$L$; $L$\,$=$\,$\{8,12,16\}$. In Fig.~\ref{fig:fig2}(a) and hereafter, increasing system sizes are denoted by circles ($L$\,$=$\,$8$), triangles ($L$\,$=$\,$12$), and stars ($L$\,$=$\,$16$). To confirm the second-order nature of this transition, we attempt a universal scaling collapse of the order parameter near the critical point according to $F$\,$=$\,$L^{-\beta/\nu}f((\Delta-\Delta_c)L^{1/\nu})$---where $\nu$ and $\beta$ are the correlation length and magnetization critical exponents, respectively---while simultaneously scaling the temperature as $T$\,$\sim$\,$1/L$, in accordance with the $z$\,$=$\,$1$ dynamical critical exponent of the underlying CFT~\cite{sachdev2011quantum}. We obtain good data collapse [Fig.~\ref{fig:fig2}(a); inset] and extract the exponents  $\nu$\,$\approx$\,$0.632$, $\beta$\,$\approx$\,$0.29$, consistent with the ($2$\,$+$\,$1$)D Ising universality class~\cite{el2014solving}. A detailed summary of the exponent-extraction procedure is presented in Appendix~\ref{app:critexp}.
	
    Now, we turn our attention to the transition between the checkerboard and the striated phase. Even though the two phases break different symmetries, the latter effectively breaks a second $\mathbb{Z}_2$ symmetry \textit{on top} of the one already broken by the former. Therefore, a second-order transition is still generically allowed. Indeed, on tuning our system across the phase boundary at a constant detuning $\Delta/\Omega$\,$=$\,$2.6$, we observe a smooth behavior of the Rydberg density, and the order parameter collapses under universal scaling with the exponents $\nu$\,$\approx$\,$0.612$, $\beta$\,$\approx$\,$0.314$ [see Fig.~\ref{fig:fig2}(b)], again consistent with the ($2$\,$+$\,$1$)D Ising universality class.
        
    Next, we investigate the transition from the disordered to the striated phase, keeping the blockade radius fixed at $R_b$\,$=$\,$1.45$. In Fig.~\ref{fig:fig2}(c,d), we show the order parameter $F(0,\pi)$ and the Rydberg density $n$ across this boundary. Both observables feature a sharp jump at the critical point, which converges towards a step function for larger system sizes, indicating a potential first-order transition. To further corroborate this claim, we plot a histogram of Rydberg densities obtained during the QMC sampling, from multiple random seeds, at the transition point. This shows a clear double-peaked distribution, conveying a coexistence of the two phases near the transition. These features, together with the lack of universal scaling, strongly suggest that this is indeed a first-order QPT.
    
\begin{figure}
	    \centering
	    \includegraphics[width=\linewidth]{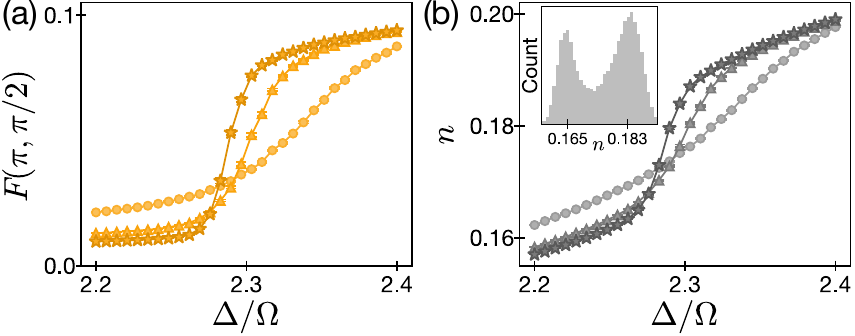}
	    \caption{Transition between the disordered and star phases. (a)--(b) Both the order parameter and the Rydberg density converge to sharp step functions for increasing system sizes. (b) Double-peaked distribution of density measurements at the phase boundary indicates a first-order transition.}
	    \label{fig:fig4}
	\end{figure}

    Finally, we simulate the transition from the disordered phase into the star phase at  $R_b$\,$=$\,$1.7$. In Fig.~\ref{fig:fig4}, we see a behavior similar to the transition into the striated phase: both the order parameter and the Rydberg density seem to converge towards a sharp step-like function. We again plot the histogram of the density sampling [Fig.~\ref{fig:fig4}(b); inset] and observe a clear double-peaked distribution, reflecting the first-order nature of the transition. 
    
    The star and striated phases break different symmetries; thus, in the absence of any exotic mechanisms such as deconfined quantum criticality~\cite{Senthil.2004}, the QPT between them must be first-order. We study this transition numerically and find that our QMC algorithm exhibits diverging equilibration times near the phase boundary, indicating the coexistence of two incompatible symmetry-breaking patterns. We employ a seeding procedure where the simulation first equilibrates deep within one of the phases and is then slowly driven towards the transition point. This method results in a convergence to a sharp jump indicating the first-order transition, as expected. Further details on extracting the phase boundary are presented in Appendix~\ref{app:seeding}.

    In summary, two of the three possible transitions between the disordered, checkerboard, and striated phases prove to be second-order while the third is seen to be first-order. The star phase is connected to the neighboring phases considered here solely through first-order QPTs. To understand the origins of these differences, we now analyze the different QPTs using the framework of LGW theory.
    \section{Field-theoretic description}\label{sec:lgw}
    
    Having identified the locations and orders of the various quantum phase transitions from numerical simulations,  we now turn to their theoretical descriptions.
    More specifically, we construct effective LGW theories~\cite{landau1937theory} to describe the nature of the phase transitions observed in the square-lattice Rydberg atom arrays. Here, we present the main results while the detailed analysis is summarized in Appendix~\ref{app:lgw}. 
    
    Focusing on the long-wavelength and low-energy physics, the key tenet of LGW theory is the ``soft-spin'' approximation~\cite{sachdev2011quantum}, which promotes the discrete local density $n_i$ at each site $i$ to a coarse-grained continuous density field, $\rho(\vect{r})$, that can be expanded in the basis set of the real-space eigenfunctions of the $N$ lowest-energy modes as
\begin{alignat}{1}
\label{eq:phi}
\rho (\vect{r}) = \mathrm{Re} \left( \sum_{n=1}^{N} \psi^{\phantom{\dagger}}_n \,\mathrm{e}^{i \vect{k}^{}_n\cdot \vect{r}}  \right),
\end{alignat}
where $\psi^{}_n$\,$\in$\,$\mathbb{C}$ is the order parameter corresponding to the $n$-th mode. The momentum-space positions of these soft modes can be identified from the peaks in the Fourier spectra of the real-space density-wave profiles. The Landau functional is then given by all homogeneous quartic polynomials in the amplitudes $\psi^{}_n$ which are invariant under the symmetry transformations of the underlying square lattice \cite{balents2005putting, huh2011vison, grass2011quantum, hwang2015z}. 
In this spirit, the Rydberg density in or around the striated phase can be expanded in terms of three \textit{real} fields $\Psi^{}_1$, $\Psi^{}_2$, and $\Phi$ as
$\rho(\vect{r})$\,$=$\,$\Psi_1^{}\, \mathrm{e}^{i \,(\pi,0)\cdot\vect{r}}$\,$ +$\,$ \Psi_2^{}\, \mathrm{e}^{i \,(0, \pi)\cdot\vect{r}}$\,$ + $\,$\Phi^{}\, \mathrm{e}^{i \,(\pi, \pi)\cdot\vect{r}}
$; note that this set already includes the $(\pi, \pi)$ Fourier peak of the checkerboard phase as well. The symmetry properties of the order parameters show that $\Phi$\,$ \sim$\,$\Psi_1 \Psi_2$ and the interplay between the three bears interesting consequences for the Landau theory.
Up to quartic order, the most general effective Hamiltonian is
\begin{alignat}{1}
&\mathcal{H}^{}_{\textsc{lgw}} = 
\nonumber \int d^D \vect{x} \Bigg[ \sum_{i=1}^2 \left(\partial_\mu \Psi_i^{} \right)^2 + \left(\partial_\mu \Phi\right)^2 \\
\nonumber &+ r \left(\Psi_1^2+\Psi_2^2 \right)+ s\, \Phi^2 + g\, \Psi^{}_1\Psi^{}_2\Phi_1 +u^{}_1 \left(\Psi_1^2+\Psi_2^2 \right)^2 \\ 
&+ u^{}_2 \Phi^4 + v\,  \Psi_1^2\,  \Psi_2^2 +w\, \Phi^2\left(\Psi_1^2+\Psi_2^2 \right) \Bigg],\label{eq:LFull}
\end{alignat}
where $\vect{x}$ denotes a $D$\,$=$\,$2+1$-dimensional spacetime coordinate, and we have suppressed the explicit dependence of the $\Psi$ and $\Phi$ fields on the continuum position $\vect{r}$. We have also rescaled all the field variables so as to make the coefficients of the gradient terms equal to unity. The parameters $r$ and $s$ tune the system across the various phase transitions. The cubic coupling $g$ is allowed by all the symmetries and plays an essential role in establishing the important features of the phase diagram.

\begin{figure}
	    \centering
	    \includegraphics[width=\linewidth]{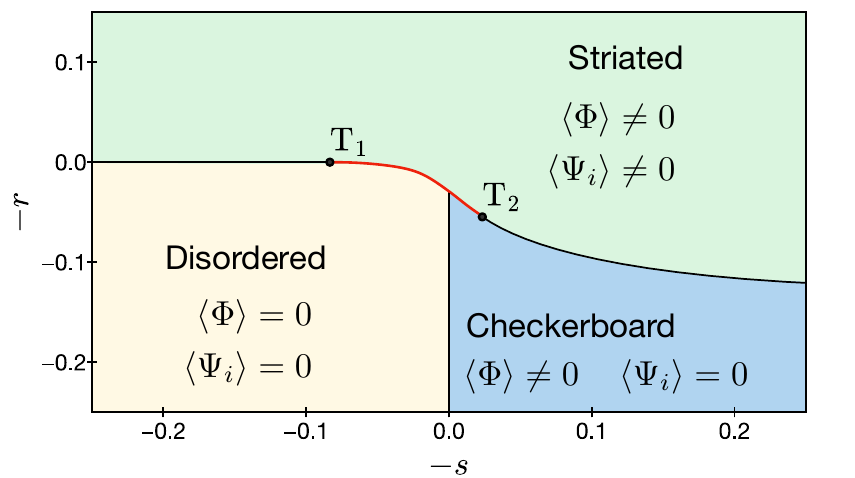}
	    \caption{Mean-field phase diagram of the low-energy Landau theory in Eq.~\eqref{eq:LFull}, illustrating the disordered, checkerboard, and striated phases as well as the fields condensed in each. The black and red lines represent second-order and first-order QPTs, respectively. The black dots mark the two tricritical points $\mbox{T}_1$ and $\mbox{T}_2$. The numerical minimization was performed taking $g$\,$=$\,$-u_1$\,$=$\,$v$\,$=$\,$-1$, $u_2$\,$=$\,$0.75$, and $w$\,$=$\,$0.5$ in  Eq.~\eqref{eq:LFull}.}
	    \label{fig:fig3}
	\end{figure}

A mean-field analysis of $\mathcal{H}_{\textsc{lgw}}$ leads to the phase diagram in Fig.~\ref{fig:fig3}, yielding (i) the trivial disordered phase, where no lattice symmetry is broken and $\langle \Phi\rangle$\,$=$\,$\langle \Psi_i\rangle$\,$=$\,$0$, (ii) the checkerboard phase, where only the $\Phi$ field is condensed, i.e., $\langle \Phi\rangle$\,$\ne$\,$0$,\,$\langle \Psi_i\rangle$\,$=$\,$0$, and (iii) the striated phase, where both the order parameters are
nonzero, so $\langle \Phi\rangle$\,$\ne$\,$0$,\,$\langle \Psi_i\rangle$\,$\ne$\,$0$. Although there can be a second-order QPT between any two of these three phases, the presence of the cubic term in Eq.~\eqref{eq:LFull} implies the existence of a line of first-order transitions close to the origin $r$\,$=$\,$s$\,$=$\,$0$~\cite{PhysRevB.64.184510, sachdev2002ground, shackleton2021deconfined}. This line terminates in two tricritical points (labeled $\mbox{T}_1$ and $\mbox{T}_2$ in Fig.~\ref{fig:fig3}), at which the coefficients of both the quadratic and quartic terms of the effective theory vanish; the theory is then controlled by its sextic term [not shown in Eq.~\eqref{eq:LFull}]. 

 Let us now address the role of the fluctuations neglected in the mean-field calculation so far by considering a more careful RG analysis. 
 For the purpose of describing the QPT from the disordered to the striated phase, the most general Hamiltonian density consistent with square-lattice symmetries can be written as
\begin{align}
\frac{1}{2} \sum_{i=1}^2\left\{\left(\partial^{}_\mu \Psi^{}_i\right)^2 +  r\, \Psi_i^2 \right\} \nonumber+ \frac{1}{4!}\, \sum_{i=1}^2 \left(u_0^{}+v_0^{} \delta^{}_{ij} \right) \Psi_i^2 \Psi_j^2.
\end{align}
This theory is known to have four RG fixed points (FPs)~\cite{PhysRevB.8.4270, aharony1976phase}, but in $D$\,$=$\,$3$, only the O($2$)-symmetric FP describes the generic critical behavior of the system. However, there is an extended region in the $(u^{}_0, v^{}_0)$ parameter plane, defined by the wedge $\{(u^{}_0, v^{}_0)$\,$\vert$\,$-v^{}_0/2$\,$<$\,$u^{}_0$\,$<$\,$0\}$, from which the FP is \textit{inaccessible}, thus rendering the transition first-order~\cite{PhysRevB.23.3943}.

Therefore, there are two possible mechanisms which could lead to the first-order transition between the disordered and striated phases observed numerically. Firstly, we could have a scenario where the star phase sets in before the tricritical point T$_1$ in Fig.~\ref{fig:fig3}, ensuring that the entire line of transitions between the disordered and the striated phase remains first-order. Alternatively, if the parameters of our theory place us in the abovementioned swath of $(u^{}_0, v^{}_0)$ space, one could have a fluctuation-induced first-order transition due to the inaccessibility of the relevant FP.

On the other hand, the $\mathbb{Z}_2$ symmetry-breaking on going from the checkerboard to the striated phase is described by the standard $\Phi^4$ field theory, where, for $D$\,$<$\,$4$, the physics of the critical point is given by the celebrated Wilson-Fisher FP \cite{PhysRevB.4.3174, PhysRevLett.28.240, PhysRevLett.28.548}. Therefore, any second-order QPT between these two phases must be in the universality class of the ($2$\,$+$\,$1$)D Ising model.
	
Having detailed the LGW theory of QPTs in the lower part of the phase diagram, we now turn to the star phase.
The QPT from the disordered to the star phase involves four real fields and is described by a three-dimensional O($4$)-symmetric vector model with anisotropic perturbations~\cite{Vicari2011}. The effective Hamiltonian, consistent with all symmetries, is (up to quartic order)
\begin{align}
\label{eq:Hstar}
{\cal H}_\phi^{} = &\int d^D \vect{x} 
\Bigg[ \sum_{i,a}^{2}\frac{1}{2}
\left\{ (\partial^{}_\mu \phi^{}_{a,i})^2 + r \phi_{a,i}^2 \right\}\nn\\
&+  \sum_{ij,ab}^{2} \frac{1}{4!}\left( u^{}_0 + v^{}_0 \delta^{}_{ij} + w^{}_0 \delta_{ij}\delta^{}_{ab} \right)
\phi^2_{a,i} \phi^2_{b,j} \Bigg],
\end{align}
where the coefficients $u_0$,$v_0$,$w_0$ must satisfy
\begin{equation}
u^{}_0 > 0 \mbox{ and }
\begin{cases}
- \left(u^{}_0 + v^{}_0\right) <{\displaystyle\frac{w^{}_0}{2}}<- v^{}_0, &\mbox{ for  } w^{}_0 >0
\\
-\left(u^{}_0 + v^{}_0\right) <w^{}_0<- v^{}_0, &\mbox{ for  } w^{}_0 <0
\end{cases}
\label{eq:region}
\end{equation}
to ensure the stability of the theory and the appropriate condensation of fields in the star phase.
Within the framework of the $\varepsilon$ expansion, this so-called tetragonal theory has eight FPs~\cite{mukamel1975physical,mukamel1975epsilon,mukamel1976physical}. However, a careful analysis shows that there are only three possible stable FPs~\cite{Pelissetto2000}: the cubic one in the $v_0^{}$\,$=$\,$0$ plane as well as its symmetric counterpart, and the XY FP with $u_0^{}$\,$=$\,$w_0^{}$\,$=$\,$0$. The cubic FP, which is stable in the $v_0^{}$\,$=$\,$0$ subspace, is unstable with respect to the quartic interaction associated with the coupling $v_0^{}$; similar considerations apply to the other cubic FP which is stable in the $v^{}_0$\,$+$\,$(3/2)w^{}_0$\,$ =$\,$ 0$ plane~\cite{PhysRevE.64.047104}. The XY FP is stable on the $u_0^{}$\,$=$\,$0$ plane and while its behavior in the full parameter space has been a subject of much debate~\cite{PhysRevB.32.4763,PhysRevB.57.3562,PhysRevB.57.5704,mudrov2001critical,PhysRevB.64.214423}, recent six-loop fixed-dimension expansion calculations~\cite{Pelissetto2000} have confirmed its global stability [including on the $w^{}_0$\,$=$\,$0$ plane in Fig.~\ref{fig:Flow2}(c)]. Thus, systems described by the tetragonal Hamiltonian are generically expected to demonstrate XY critical behavior. 

Crucially, the XY FP is rendered inaccessible from the allowed region in the parameter space of our effective theories given by Eq.~\eqref{eq:region} as shown in Appendix~\ref{app:lgw}. Therefore, the QPT between the disordered and star phases is a \emph{fluctuation-induced} first-order transition. 
	\section{Boundary criticality}\label{sec:boundary}
	The notion of a QPT is formally defined in the thermodynamic limit. In practice though, one always considers a finite system, in which case the boundaries may have important effects on the critical behavior~\cite{diehl1997theory}. Concentrating on the disordered--striated transition in an array endowed with OBC, here, we address this issue in the context of the experimentally realizable Rydberg phase diagram. 
	
	\begin{figure}
	    \centering
	    \includegraphics[width=\linewidth]{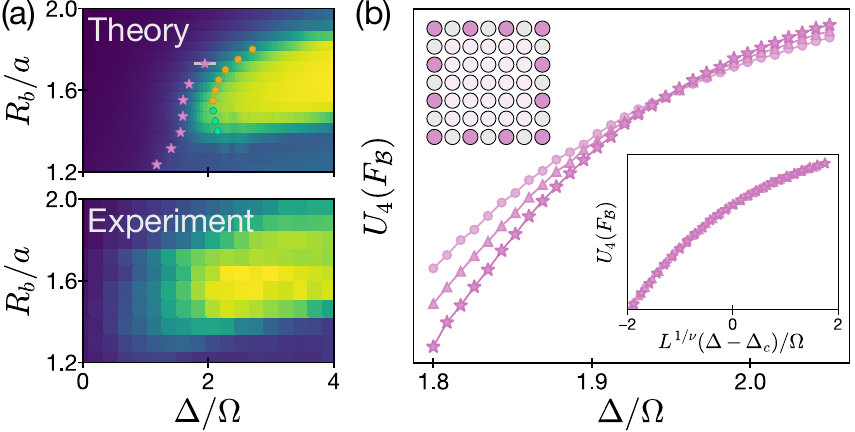}
	    \caption{Boundary phase transition and its implications for experiments,
	    (a) Comparison of the striated order parameter (color map) obtained from our QMC simulation with OBC against that of the experiment in Ref.~\onlinecite{Ebadi.2021}, showing good agreement. The boundary orders first (lower $\Delta/\Omega$, pink stars) and strongly influences the extent of the striated phase, which is extended to a wider range of parameters compared to the bulk behavior on a torus. Orange and green dots denote phase transitions in a system with PBC [Fig.~\ref{fig:fig1}]. (b) Binder ratio of the boundary order parameter across the transition marked in gray in (a). The insets show the boundary ordering (upper left) and universal collapse with 1D exponents (lower right).}
	    \label{fig:fig5}
	\end{figure}
	
	Remarkably, we notice that the boundary itself undergoes a second-order QPT \textit{before} the bulk in the thermodynamic limit [top panel of Fig.~\ref{fig:fig5}(a)]. Intuitively, atoms at the boundary have fewer neighbours (and interactions), so it is easier for them to order. For a finite-size system, this surface order shifts the onset of bulk ordering  towards smaller $\Delta/\Omega$  because the boundary ``seeds'' the interior of the system. Such an onset of bulk order in the presence of established boundary order defines an \emph{extraordinary} boundary universality class~\cite{metlitski2020boundary}. Furthermore, since the $\mathbb{Z}_2$ ordering at the boundary is compatible with the checkerboard and striated phases---but not with the star---the striated phase is significantly expanded, relative to a system with PBC, at the cost of the shrunken star phase.
	
	Next, we determine the critical exponents of the boundary transition by calculating the boundary order parameter $F_{\mathcal{B}}$\,$\equiv$\,$[\tilde{F}_{\mathcal B}(\pi,\pi)$\,$+$\,$\tilde{F}_{\mathcal B}(\pi,\pi)]/2$ with
	\begin{equation}
	    \tilde{F}_{\mathcal B}(k_x,k_y) \equiv \frac{1}{N_{\mathcal B}} \hspace*{-.1cm}\sum_{j\in{\rm Boundary}}\hspace*{-.35cm} n_j \,\exp \left[i(k_x,k_y)\cdot(x_j,y_j)\right],
	    \label{eq:Fb}
	\end{equation}
$N_{\mathcal B}$ being the number of atoms along the boundary.
 In Fig.~\ref{fig:fig5}(b), we present the Binder ratio of $F_{\mathcal{B}}$ while the inset illustrates its universal collapse with $\nu$\,$=$\,$1.004(8)$, consistent with a $(1$\,$+$\,$1)$D Ising transition. 
Signatures of this boundary ordering have been also obtained from experimental data using machine learning in a complementary work~\cite{Kim.2021}.

The ordering of the boundary has direct implications for quasi-adiabatic state preparation across a critical point. According to the Kibble-Zurek~\cite{Keesling.2019,kibble1976topology,kibble1980some,zurek1985cosmological,zurek1993cosmic,zurek1996cosmological} mechanism, the scaling of the density of defects in the resultant phase after a quench is governed by the universal static critical exponents of the QPT. Since $(1$\,$+$\,$1$)D and $(2$\,$+$\,$1$)D Ising universality classes exhibit scaling behavior with different critical exponents, the efficiency of adiabatic state preparation may depend on the interplay between the boundary and the bulk. 
  For example,  during  dynamical preparation, the system could undergo a \emph{cascade} of boundary transitions propagating inwards, effectively masking
 the bulk first-order transition.

Apart from  understanding the existing experimental observations, an important open question is whether one can leverage the boundary ordering to facilitate improved  preparation of bulk-ordered states in larger systems.
 Detailed understanding of these processes requires careful theoretical and experimental studies of real-time quantum dynamics of large systems.
 \section{Discussion and outlook}\label{sec:outlook}
	The results in this work demonstrate the emergence of both first- and second-order QPTs in a square-lattice Rydberg array with periodic boundary conditions and present their unified field-theoretic description. Furthermore, we identified the crucial role of boundary ordering in systems with OBC, which allows one to adiabatically access compatible phases that are otherwise hidden behind a first-order transition line.
	Our findings suggest that the  boundary plays a major role in understanding experimental results. 

These studies can be extended along several directions. First, we note that at stronger interactions, other phases are known to emerge for the system at hand \cite{Samajdar_2020}. The classification and theoretical study of those orders, as well as different lattice geometries~\cite{Samajdar.2021,Verresen.2020,Ebadi.2021} or interparticle interactions, is an interesting direction for future work.
Second, the application of the 
phenomena discussed in this work to disordered systems warrants a separate investigation. Understanding the role of boundary ordering and quasi-adiabatic state preparation in these settings is a key factor in predicting practical performance and potential quantum advantage  of  near-term quantum simulators for optimization problems such as the Maximum Independent Set~\cite{KimWires.2021,MIS}.

	\begin{acknowledgments}
	We thank M.~Cain, E.-A.~Kim, C.~Miles, H.~Pichler, R.~Sahay, H.~Shackleton, E.~Vicari, and especially the team of D.~Bluvstein, S.~Ebadi, H.~Levine, A.~Omran, A.~Keesling, G.~Semeghini, and T.~T.~Wang for useful discussions. R.S. and S.S. are supported by the U.S. Department of Energy under Grant DE-SC0019030. R.M. is supported by NSERC, the Canada Research Chair program, and the Perimeter Institute for Theoretical Physics. Research at Perimeter Institute is supported in part by the Government of Canada through the Department of Innovation, Science and Economic Development Canada and by the Province of Ontario through the Ministry of Economic Development, Job Creation and Trade. M.D.L. is supported by the U.S.~Department of Energy under Grant $\mbox{DE-SC0021013}$,  the Harvard--MIT Center for Ultracold Atoms, the Office of Naval Research, and the Vannevar Bush Faculty Fellowship. The computations in this paper were run on the FASRC Cannon cluster supported by the FAS Division of Science Research Computing Group at Harvard University. 
	\end{acknowledgments}
	
	\appendix
	
	\section{Quantum Monte Carlo simulation}\label{app:qmc}
Our algorithm is based on the work in Refs.~\onlinecite{Nakamura.2003,Nakamura_2008}; here, we briefly summarize it for completeness. We found that, for the Hamiltonian in Eq.~(1) of the main text, this QMC method performs better than the conventional stochastic series expansion algorithm with cluster updates~\cite{Sandvik.2003}, which we attribute to the presence of strong nearest-neighbor blockade interactions that violate the Ising symmetry.
     Our QMC scheme operates at a finite temperature $T$, and, in order to access properties of the ground state, we work at sufficiently low temperatures, e.g. $T/\Omega$\,$\sim$\,$0.01$ and $0.02$. 
     
     We write our $d$-dimensional system's Hamiltonian as $\hat{H}=\hat{H}_0+\hat{H}_1$, with
\begin{equation}
    H = \underbrace{\sum_{i<j} V^{}_{ij}\hat{n}^{}_i \hat{n}^{}_j -\Delta \sum_{i}\hat{n}^{}_i}_{\hat{H}_0} \underbrace{-\frac{\Omega}{2}\sum_{i} \hat{\sigma}_i^x}_{\hat{H}_1},
\end{equation}
where $\hat{H}_0$ denotes the part diagonal in the $z$-basis and $\hat{H}_1$ is the off-diagonal term proportional to $\hat{\sigma}^x$. In the discrete imaginary-time representation (with $N_I$ sites in the imaginary-time direction), the partition function is
\begin{equation}\label{seq:Z}
    Z = \tr[e^{-\beta H}] = \lim_{N_I\to\infty}\sum_{\vec{\alpha}} \prod_{a=0}^{N_I-1} \bra{\alpha^{}_{a+1}}e^{-\tau \hat{H}_{\rm 0}}e^{-\tau \hat{H}_{\rm 1}}\ket{\alpha^{}_{a}},
\end{equation}
where $\vec{\alpha}={\alpha_0,\alpha_1,...,\alpha_{N_I}}$, $\alpha_i$ is a state in the computational ($z$) basis with $\alpha_{N_I}=\alpha_0$ due to the periodicity induced by the trace, and $\tau = \beta/M$ is assumed to be very small such that the above Suzuki-Trotter decomposition holds. The partition function in Eq.~\eqref{seq:Z} can be recast as a partition function of a $(d+1)$-dimensional classical model, with the index $a$ labeling the additional dimension:
\begin{equation}
    Z^{}_{\rm cl} = \sum_{\alpha_{\rm cl}} \braket{\alpha^{}_{\rm cl}|e^{-\beta H_{\rm cl}}|\alpha^{}_{\rm cl}},
\end{equation}
where $\alpha_{\rm cl}$ are computational basis states of $N_I\times {N_a}$ spins and 
\begin{align}
    H^{}_{\rm cl} = \sum_{a=0}^{N_I-1}\Bigg(\sum_{i<j}^N \frac{V^{}_{ij}}{N^{}_I}\hat{n}^{}_{a,i} \hat{n}^{}_{a,j} -\frac{\Delta}{N^{}_I} \sum^N_{i}\hat{n}^{}_i\nn\\-\frac{\ln\coth\left(\frac{\beta\, \Omega}{2N_I}\right)}{2\beta}\sum^N_{i}\hat{\sigma}^{z}_{a,i}\hat{\sigma}^z_{a+1,i}\Bigg),
\end{align}
where $a$ labels the imaginary-time direction. This Hamiltonian is diagonal in the computational basis. In order to avoid errors stemming from the Suzuki-Trotter decomposition, one has to use large values of $M$, which makes the simulation inefficient. 

\renewcommand{\arraystretch}{1.2}
\begin{table*}[tb]
    \centering
    Fitting of the order parameter $F(\vect{k})$
\\
    \begin{tabular}{l c c c c c c c c }
    \hline
    \hline
         Transition & $L_{\rm min}$  & $\beta$ & $a_0$ & $a_1\times 10^3$ & $a_2\times 10^3$ & $a_3\times 10^3$ & $a_4\times 10^3$ \\
         \hline
        Disordered$\,\leftrightarrow\,$checkerboard & 12 &0.291(1) &  0.309(1) & 67.9(3) & 5.6(1) & $-$0.96(2) & $-$0.035(3)\\
        
        Checkerboard$\,\leftrightarrow\,$striated & 12 & 0.314(1) & 0.225(1) & 27.4(1) & 20.3(5) & $-$3.8(8) & $-2.7(8)$  \\
        \hline
        \hline
    \end{tabular}\\
    \vspace{20pt}
    Fitting of the Binder ratio $U_4$
\\
    \begin{tabular}{l c c c c c c c c }
    \hline
    \hline
         Transition & $L_{\rm min}$ & $g_c$ & $\nu$ & $b_0$ & $b_1\times 10^3$ & $b_2\times 10^3$ & $b_3\times 10^3$ & $b_4\times 10^3$\\
         \hline
        Disordered$\,\leftrightarrow\,$checkerboard & 12 & 1.0959(1) & 0.632(5) & 0.7048(9)  & 122(4) & $-$9.6(6)  & -2.5(2) & 0.34(5) \\
        Checkerboard$\,\leftrightarrow\,$striated & 12 & 1.38000(1) & 0.612(6) & 0.591(1) & 740(34) & 5(21) & $-$673(82) & 054(157) \\
        \hline
        \hline
    \end{tabular}
    \caption{Explicit values of parameters obtained for the fitting ans\"atze in Eqs.~\eqref{seq:U4} and \eqref{seq:F}. These values correspond to the data shown in Fig.~\ref{sfig:exponent}. The data for system size $L=8$ exhibits strong finite-size effects, so we use system sizes $12,16,20$ for extracting the exponents.\\}
    \label{stab:exp}
\end{table*}

\begin{figure*}[tb]
    \centering
    \includegraphics[width=1.0\linewidth]{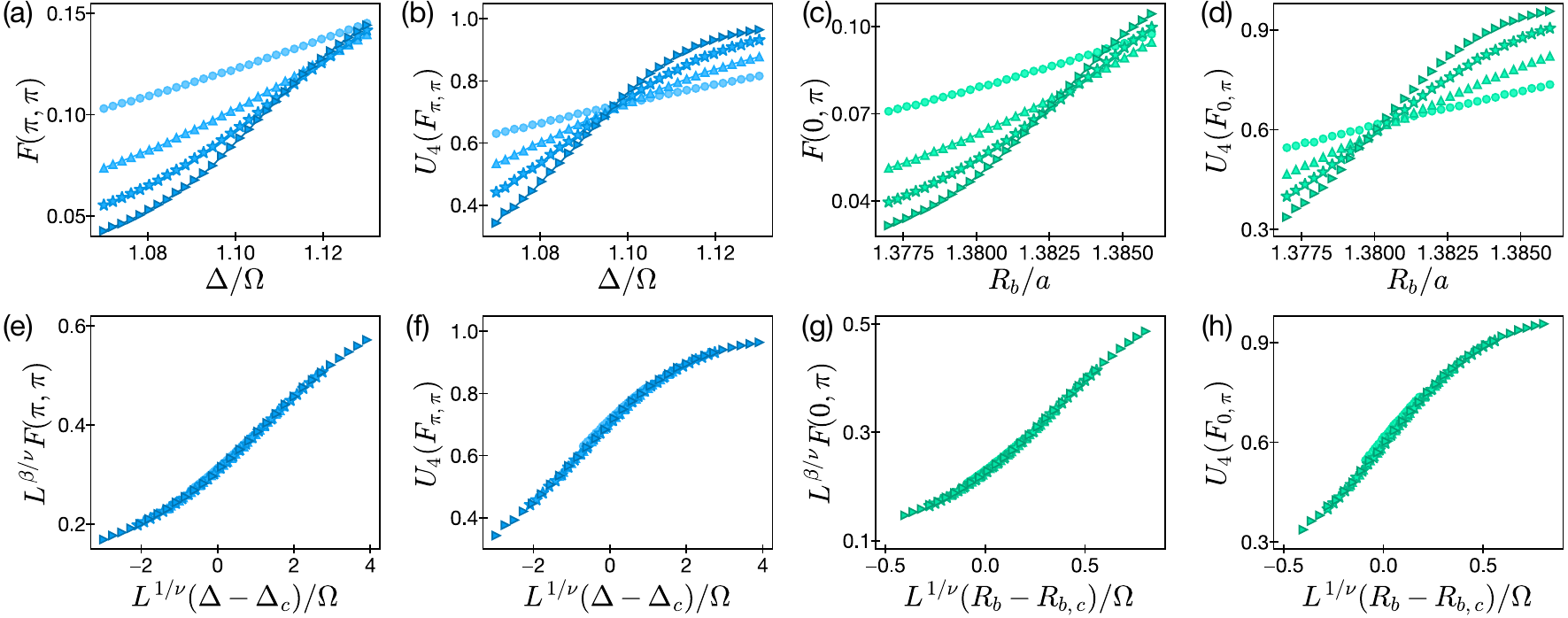}
    \caption{Extraction of critical exponents. System sizes are (in increasing color intensity) $8\times8$, $12\times12$, $16\times16$, and $20\times 20$. (a--b) The order parameter and the Binder ratio for the transition to the checkerboard phase. (c--d) The order parameter and the Binder ratio for the transition to the striated phase from the checkerboard phase. (e--h) Corresponding curves in the top row exhibiting data collapse with the extracted critical exponents.}
    \label{sfig:exponent}
\end{figure*}

To amend this issue, Refs.~\onlinecite{Nakamura.2003,Nakamura_2008} go to the limit of $N_I\to\infty$ directly. This can be achieved by keeping track of domain walls in the imaginary-time direction, rather than individual spins. Assuming that there are reasonably many clusters, this approach should reduce the computational complexity by a large factor. Fortunately, cluster lengths obey the Poisson distribution~\cite{Evertz.2003}; therefore, in each update step, we can sample potential domain wall positions accordingly and attempt to flip whole clusters using the usual importance sampling. This scheme is therefore nonlocal in the imaginary-time direction (cluster update) but local in space (local update). 

We note that an independent work in Ref.~\onlinecite{Merali.2021} developed a similar method, which is local in space and nonlocal in imaginary time, based on the Stochastic Series Expansion (SSE) approach. In certain parameter regimes,~\citet{Merali.2021} observe better performance using this ``line method'' compared to the usual cluster updates for SSE approach.

\section{Extraction of critical exponents}\label{app:critexp}

\begin{figure*}[ht]
    \centering
    \includegraphics[width=\linewidth]{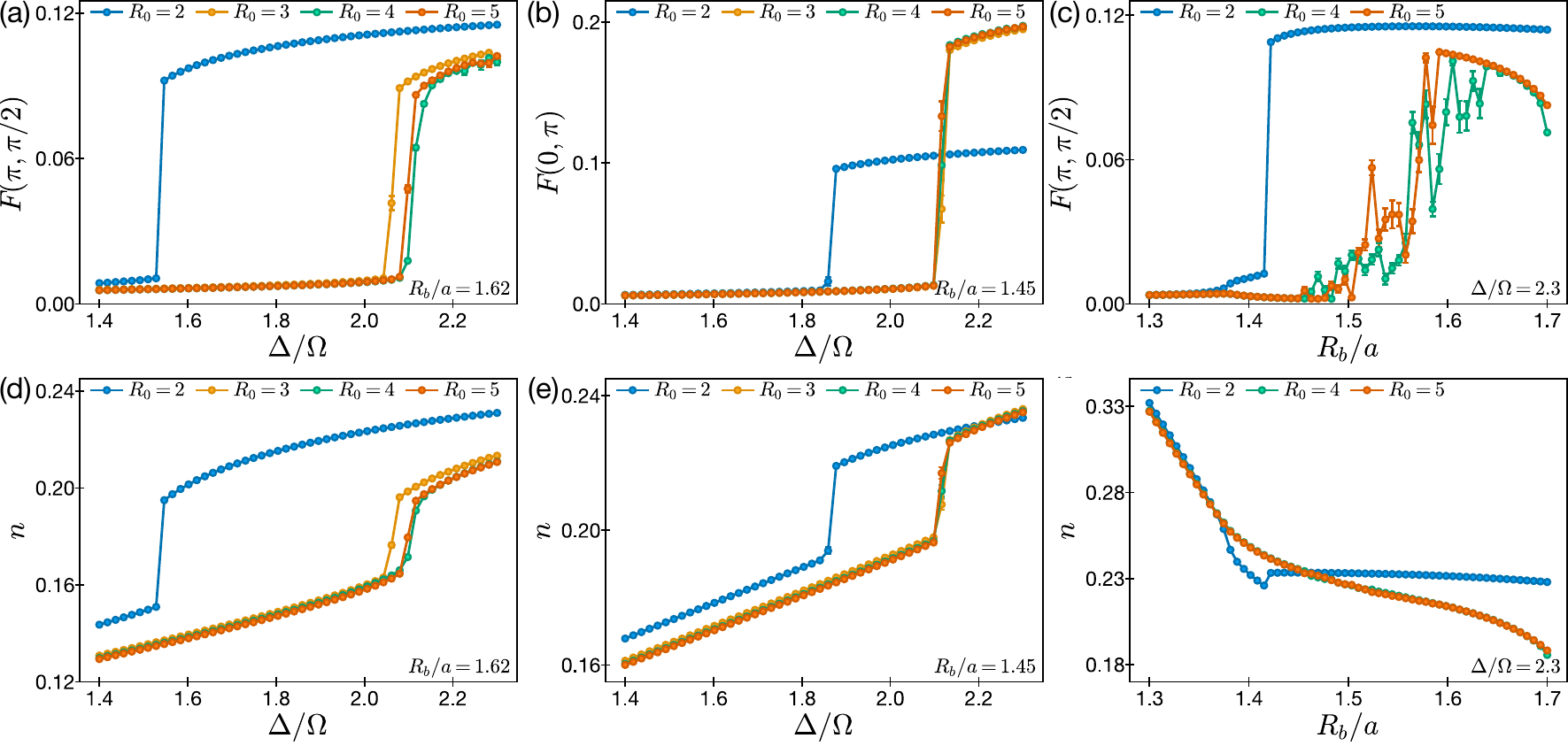}
    \caption{Effect of truncating interactions at different distances $R_0$ for a system of size $16\times 16$. (a,d) Transition to the star phase at $R_b=1.62$. The slight discrepancy between $R_0=4$ and $R_0=5$ suggests that the transition can be even sharper at $R_0=5$, at this value of the blockade radius. (b,e) Transitions at $R_b=1.45$. For $R_0=2$, the transition is to the star phase, while for $R_0\in \{3,4,5\}$, one obtains a striated phase. (c,f) Transition from the checkerboard to the striated phase for a fixed $\Omega/\Delta=2.3$. Taking $R_0 = 5$ slightly stabilizes the star phase but does not resolve the difficulty of simulating the interface between these two phases.}
    \label{sfig:cutoff}
\end{figure*}

For second-order phase transitions, we assume a dynamical critical exponent of $z$\,$=$\,$1$ and thus scale the temperature with the linear system size as $T\sim 1/L$. We are interested in the $\nu$ (correlation length) and $\beta$ (magnetization) exponents, which we will extract from the Binder ratio $U_4$ and the order parameter $F$, respectively. To this end, near a critical point $g_c$, we consider a universal scaling form of the Binder ratio as
\begin{align}
    U^{}_4 &= f_1[(g-g_c) L^{1/\nu}] \nn\\&= \sum_{k=0}^K a^{}_k (g-g^{}_c)^k L^{k/\nu}+ \mathcal{O}\Big((g-g^{}_c)^{K+1}\Big),\label{seq:U4}
\end{align}
where $g$ is the coupling constant being varied (in our case, $R_b$ or $\Delta$),  and $f_1$ is some universal function near the critical point.
In practice, we fit the universal function to a polynomial in the distance to the critical coupling value $(g-g_c)$ to  the $K^{\rm th}$ order with coefficients $a_k$. If the fitting procedure is stable (i.e., there is universal scaling), it should be possible to truncate the expansion at a reasonably small $K$ when close enough to the critical point. This is visible in the relatively small extracted values of $a_k$ for larger $k$. We set $K=4$ in this work and found it to be sufficient.

Similarly, we assume the universal form of the order parameter:
\begin{align}
    F 
    &= L^{-\beta/\nu}f_2[(g-g^{}_c) L^{1/\nu}] \nn\\
    &= \sum_{k=0}^K b^{}_k (g-g^{}_c)^k L^{(k-\beta)/\nu}+ \mathcal{O}\Big((g-g^{}_c)^{K+1}\Big).\label{seq:F}
\end{align}
To extract critical exponents, we calculate  the order parameter and its Binder ratio close to the critical point for different system sizes. Then, we fit the data to the ansatz of Eq.~\eqref{seq:U4} to obtain $\nu$ and subsequently use that value to fit $\beta$ from Eq.~\eqref{seq:F}. Note that the error estimates in Table~\ref{stab:exp} are results of the fitting procedure and do not include systematic (e.g., finite-size) errors. We attribute the discrepancy between the obtained exponents and those expected for the $(2+1)$D Ising universality class to such finite-size effects. Moreover, these effects seem to be more pronounced at the checkerboard--striated transition as seen in Fig.~\ref{sfig:exponent}(d), where the Binder ratios' crossing points for increasing system sizes drift significantly, compared to Fig.~\ref{sfig:exponent}(b).

\section{Cutoff dependence of the phase diagram}\label{app:cutoff}
The interaction that we use to study Eq.~(1) of the main text is truncated at a finite distance: $V(\vect{R})=\Omega\left(R_b^6/\abs{\vect{R}}^6\right)\Theta(R_0-\abs{\vect{R}})$, where $\Theta(x)$ is the Heaviside step function with $\Theta(0)=1$.  In this work, we assumed $R_0 = 4$, in units where we set the lattice spacing to unity ($a=1$). This means that a single atom can interact with up to 48 other atoms.
In Fig.~\ref{sfig:cutoff}, we present various order parameters across transitions to the star and striated phases with different interaction cutoffs $R_0\in\{2,3,4,5\}$  for a $16\times16$ lattice with PBC. We observe that setting $R_0>3$ is important for recovering the detailed features of our phase diagram. For instance, taking $R_0=2$ favors the star phase, since the intra-unit-cell interactions in the star ordering are omitted; this accounts for the reduced (enhanced) extent of the striated (star) phase in Ref.~\onlinecite{Samajdar_2020} compared to our current findings. We also note that $R_0 = 3$ is not sufficient to capture the full long-tail phase diagram for the star phase [see Fig.~\ref{sfig:cutoff}(a,b)]. Noticeably, with increasing cutoff distance, the phase boundaries in  Fig.~\ref{sfig:cutoff} shift towards larger detunings and converge for $R_0\geq 4$. In Fig.~\ref{sfig:cutoff}(b), we see that including even longer-ranged tails than those assumed in this work, i.e., $R_0$\,$=$\,$5$, leads to a sharper (stronger) first-order transition to the star phase. 

\begin{figure*}[tb]
    \centering
    \includegraphics[width=\linewidth]{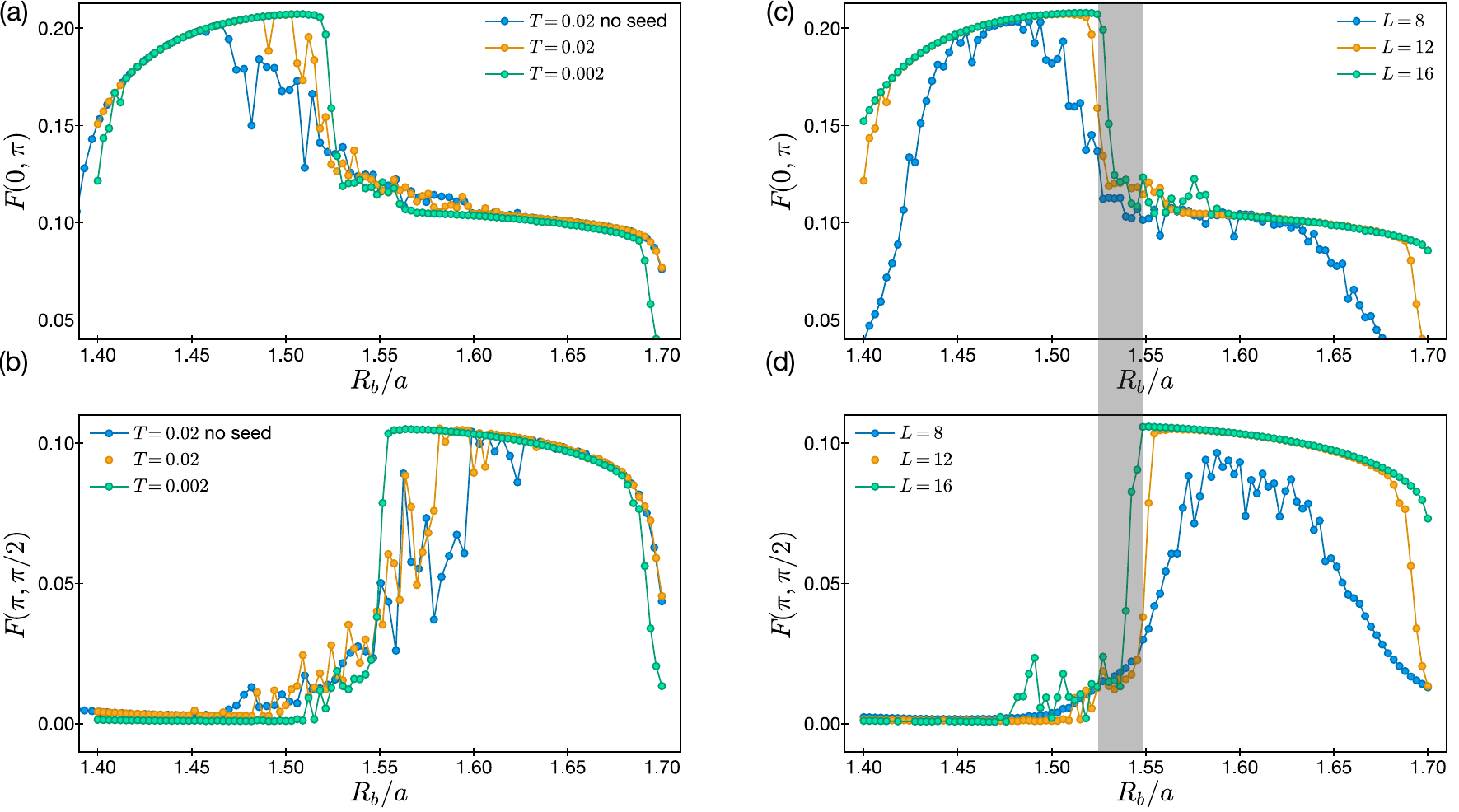}
    \caption{Seeding procedure for the interface between striated and star phases. Our QMC algorithm struggles in this regime. To estimate the phase boundary, we resort to phase seeding from both sides. (a,b) Effect of the seeding procedure and temperature change on the phase boundary at $L=12$. (c,d) Scaling with the system size at $T=0.002$. The gray region denotes our estimate of the location of the transition point.}  
    \label{sfig:seed}
\end{figure*}

\section{Seeding procedure for the striated--star transition}\label{app:seeding}
Since the star and striated phases break different symmetries, within the LGW paradigm, we expect the transition between them to be first-order. We find that our QMC algorithm struggles to converge properly near this star--striated transition. We attribute this to the local-in-space nature of our QMC update method, wherein it is necessary to reorder the whole lattice to maintain ergodicity, which is very difficult when two phases coexist.

In order to estimate the location of the transition, we perform a ``phase seeding procedure''. First, we equilibrate a QMC realization deep within one of the phases (e.g., striated), which prepares the initial ``seed'' for that particular phase. Second, we change the blockade radius inside the simulation to the desired value and then perform a normal QMC procedure (equilibration together with sampling) using the previously obtained seed instead of a completely random one. This favors the phase corresponding to the seed. Finally, we repeat this procedure starting from the other phase.

In Fig.~\ref{sfig:seed}(a), we present the effect of seeding on the behavior of the order parameters; each order parameter is seeded from its corresponding phase. We estimate the transition point as the midpoint between the peaks of the two order parameters. Upon decreasing the temperature, the effect of the seeding procedure becomes more pronounced, as the system is progressively more frozen in its initial (seeded) configuration, and the gap between the peaks becomes narrower. In Fig.~\ref{sfig:seed}(b), we show results for different system sizes at the lowest temperature considered ($T=0.002$) as well as our estimate for the transition point (grey area). We note that this position does not vary much with the temperature. 
This approach to estimating the transition point is heuristic, so we assign a rough error to the position of the critical point defined by the distance between the peaks at the lowest temperature (width of the shaded area).

\section{Details of the LGW theories}\label{app:lgw}
In order to systematically address fluctuation corrections, it is useful to regard the relevant Landau theory not as the free energy functional but rather as the Hamiltonian of a classical statistical mechanics problem in which the degrees of freedom are represented by the field(s)~\cite{sachdev2011quantum}. Landau theory would then simply follow by making the saddle-point approximation to the functional integral for the partition function. Here, we adopt this approach from the very beginning.

The elements of the space group of the square lattice include single-site translations along the $x\, (T_x$) and $y\, (T_y$) axes, reflections about the $x\, (R_x$) and $y\,(R_y$) axes, and fourfold rotations around the out-of-plane $z$ axis ($C_4$). To write down the effective Hamiltonian, such as that in Eq.~\eqref{eq:LFull} of the main text, we need to determine how the low-energy eigenmodes $\psi^{}_n$ transforms under these operations. This, in turn, follows from the transformation properties of the eigenvectors $\exp(i \vect{k}_n\cdot \vect{r})$, introduced in Eq.~\eqref{eq:phi} of the main text, as
\begin{equation*}
\hat{\mathcal{O}} \rho (\vect{r}) = \mathrm{Re}\left[ \sum_n \psi^{\phantom{\dagger}}_n\, \mathrm{e}^{i \vect{k}_n\cdot \left(\hat{\mathcal{O}} \vect{r}\right)}\right] \equiv \mathrm{Re}\left[ \sum_n \left(\hat{\mathcal{O}} \psi^{} \right)_n \mathrm{e}^{i \vect{k}_n\cdot  \vect{r}}\right].
\end{equation*}
We outline these symmetry transformations individually for each of the phases in the following.

\subsection{Checkerboard and striated phases}
The minimal set of momenta $\{ \vect{k}_n \}$ required to describe the density-wave ordering $ \rho (\vect{r})$ in the striated phase is $\{(\pi,0), (0,\pi), (\pi,\pi) \}$.
The magnetization in or around these phases can thus be expressed in terms of three \textit{real} fields $\Psi^{}_1$, $\Psi^{}_2$, and $\Phi$ as
\begin{alignat}{1}
\rho(\vect{r}) = \Psi_1^{}\, \mathrm{e}^{i \,(\pi,0)\cdot\vect{r}} + \Psi_2^{}\, \mathrm{e}^{i \,(0, \pi)\cdot\vect{r}} + \Phi^{}\, \mathrm{e}^{i \,(\pi, \pi)\cdot\vect{r}}.
\end{alignat}

In the basis $\left(\Psi^{}_1,\Psi^{}_2\right)$, the matrix representations of the symmetry transformations are:
\begin{equation}
T^{}_x = -\sigma^{}_3,\, T^{}_y = \sigma^{}_3,\, R^{}_x = R^{}_y =  \mathds{1},\,C^{}_4 =  \sigma^{}_1,
\label{eq:symm1}
\end{equation}
where $\sigma$ denotes the usual $2 \times 2$ Pauli matrices.  The field $\Phi$ transforms trivially under all symmetries except translations ($T_x, T_y$), which act as $\Phi$\,$\rightarrow$\,$-\Phi$. The Landau functional is given by all homogeneous polynomials that are
invariant under the group generated by these transformations and, up to quartic order, corresponds to Eq.~\eqref{eq:LFull} of the main text. Without loss of generality, we consider $g$\,$<$\,$0$ here. Furthermore, we need $v$\,$<$\,$0$ to ensure that both $\Psi_{1}$ and $\Psi_{2}$ condense in the ordered (striated) phase. The stability of the theory also requires $u_1$\,$>$\,$-v/4$\,$>$\,$0$, $u_2$\,$>$\,$0$, and $u_2\, (4u_1 + v)$\,$>$\,$w^2$ (assuming $w<0$).

\subsubsection{Mean-field theory for the tricritical points}
Neglecting spatial fluctuations, let us first analyze Eq.~\eqref{eq:LFull} in mean-field theory. The results of such an analysis are presented in the phase diagram of Fig.~\ref{fig:fig3} in the main text, which illustrates the checkerboard, striated, and star phases. As indicated in Fig.~\ref{fig:fig3}, there can be a second-order phase transition between any two of these three phases. From a conventional Landau theory analysis, we find that the second-order phase boundary between the disordered phase and the checkerboard phase is at $s$\,$=$\,$0$, while the line demarcating the disordered phase from the striated is at $r$\,$=$\,$0$. Finally, the second-order transition from the checkerboard to the striated is at
\begin{equation}
r = \frac{1}{2} \left(-g \sqrt{\frac{-s}{2 u^{}_2}}+ \frac{s\, w}{u^{}_2} \right).
\label{eq:2OLine}
\end{equation}
Although these three second-order lines would appear to meet at $r$\,$=$\,$s$\,$=$\,$0$, as noted by \citet{PhysRevB.64.184510} this is \textit{pre-empted} by a line of first-order transitions close to the origin. The origin of this feature can be seen by  integrating out the $\Phi$ fluctuations to derive an effective action for the $\Psi$. Doing so always induces an effective quartic term $\Psi_1^2 \Psi_2^2$ with a \textit{negative} coefficient $\sim$\,$-g^2/\lvert s \rvert$. Hence, for sufficiently small $\lvert s \rvert$, the net coefficient of $\Psi_1^2 \Psi_2^2$ always becomes negative, thus driving the transitions involving the onset of nonzero $\Psi_i$ first-order. This line of first-order transitions terminates in two tricritical points (labeled $\mbox{T}_1$ and $\mbox{T}_2$ in Fig.~\ref{fig:fig3}), at which the coefficients of both the quadratic and quartic terms of the effective theory vanish; the theory is then controlled by its sextic term [not shown in Eq.~\eqref{eq:LFull}]. The point $\mbox{T}_1$ is located at $r$\,$=$\,$ 0$, $s$\,$=$\,$g^2/(16 u_1 + 4 v)$ whereas the coordinates of $\mbox{T}_2$ can be found by solving for $s$ in the equation
\begin{equation}
\frac{g^2}{s}-4 g\, w \sqrt{\frac{2}{-s\, u^{}_2}} +8 \left(4 u + v -\frac{w^2}{u^{}_2}\right) = 0
\end{equation}
and then determining $r$ as given by Eq.~\eqref{eq:2OLine} for this value of $s$. On going across either of these two tricritical points, the change in the sign of the quartic term is responsible for the nature of the transition changing from first to second-order.

Next, we address the role of quantum fluctuations.

\subsubsection{Disordered to checkerboard}
The transition from the disordered to the checkerboard phase is characterized by the onset of a nonzero order parameter, namely, the staggered magnetization or, equivalently, the amplitude of the $(\pi, \pi)$ Fourier mode. 
Expanding in powers and gradients of this order parameter, we obtain the Hamiltonian
\begin{equation}
\mathcal{H}_\Phi^{\phantom{\dagger}} = \int d^D \vect{x} \left[ \frac{1}{2}\left\{\left(\partial^{}_\mu \Phi\right)^2 +  r\, \Phi^2 \right\}+ \frac{u^{}_0}{4!}\, \Phi^4 \right],
\label{eq:Phi4}
\end{equation}
where $\Phi$, as before, is a real, one-component field. The Hamiltonian is invariant under $\Phi$\,$ \rightarrow$\,$- \Phi$ and thus possesses a $\mathbb{Z}_2$ (Ising) symmetry.
Analyzing the renormalization-group (RG) flow of this theory, one finds that it has a Guassian fixed point (FP) at $r$\,$=$\,$u_0$\,$= 0$; however, the Gaussian FP is stable towards $u_0$ perturbations only for $D >4$. More relevantly, the RG flow has another FP at nonzero values of $r$ and $u_0$, which is the celebrated Wilson-Fisher FP~\cite{PhysRevB.4.3174, PhysRevLett.28.240,PhysRevLett.28.548} located at 
\begin{equation}
r^* = - \frac{\varepsilon}{6} + \mathcal{O}(\varepsilon^2); \quad u_0^* = \frac{2 \varepsilon}{3\, \mathcal{S}_4}+ \mathcal{O}(\varepsilon^2),
\end{equation}
where $\varepsilon \equiv 4$\,$-$\,$D$, and  the phase space factor $\mathcal{S}_d = 2/[\Gamma(d/2) (4 \pi)^{d/2}]$ denotes the surface area of a sphere in $d$ dimensions. For $D<4$, the physics of the
critical point is described by the field theory of the Wilson-Fisher FP, and the transition from the disordered to the checkerboard phase in the Rydberg system is in the universality class of the (2+1)D Ising model.

\subsubsection{Disordered to striated}
The cubic term in Eq.~\eqref{eq:LFull} of the main text implies that if two of the fields $(\Psi^{}_1,$\,$\Psi^{}_2,$\,$\Phi)$ are condensed, then so must the third~\cite{shackleton2021deconfined}.
In the striated phase, the Fourier transform of the Rydberg excitation density $\lvert n (\vect{k})\rvert$ thus exhibits peaks not only at $(\pi, 0)$ and $(0, \pi)$ but also at $(\pi, \pi)$\,$=$\,$(\pi, 0)$\,$+$\,$(0, \pi)$. For the purpose of describing the phase transition from the disordered to the striated phase, therefore, it suffices to focus on the first two momenta alone. In other words, given \textit{two} real fields $\Psi^{}_1$ and $\Psi^{}_2$,
\begin{alignat}{1}
\rho(\vect{r}) = \Psi_1^{}\, \mathrm{e}^{i \,(\pi,0)\cdot\vect{r}} + \Psi_2^{}\, \mathrm{e}^{i \,(0, \pi)\cdot\vect{r}}
\end{alignat}
correctly describes the $\mathbb{Z}_2$\,$\times$\,$\mathbb{Z}_2$ symmetry-breaking pattern of the striated phase.
Using  the matrix representations of the transformations  in Eq.~\eqref{eq:symm1}, the most general Hamiltonian consistent with square-lattice symmetries can be written as
\begin{align}
\mathcal{H}_\Psi^{\phantom{\dagger}} = \int d^D \vect{x} \Bigg[ \frac{1}{2} \sum_{i=1}^N\left\{\left(\partial^{}_\mu \Psi^{}_i\right)^2 +  r\, \Psi_i^2 \right\}\nn\\
+ \frac{1}{4!}\, \sum_{i=1}^N \left(u_0^{}+v_0^{} \delta^{}_{ij} \right) \Psi_i^2 \Psi_j^2 \Bigg],
\label{eq:HStr}
\end{align}
where $N$\,$=$\,$2$. The cubic-symmetric quartic term $\sum_i \Psi^4_i$ breaks the O($N$) invariance of the model down to a residual discrete $D_4$ symmetry. The relevance of the anisotropic perturbations can be directly understood by classifying them using irreducible representations of the O$(N)$ internal group, and computing the RG dimensions of their associated couplings~\cite{Vicari2011}.  While the coupling constants $r$,\,$u^{}_0$, and $v^{}_0$ can be varied by tuning the parameters of the microscopic Rydberg Hamiltonian, the stability of the theory requires the positivity conditions 
\begin{equation}
u^{}_0 + v^{}_0 > 0 \quad \mbox{and} \quad  N u^{}_0 + v^{}_0 > 0 
\label{eq:stability}
\end{equation}
to be satisfied. Rewriting the quartic term as $[(u^{}_0+v^{}_0)(\Psi^2_1 + \Psi_2^2)^2 - 2\, v^{}_0 \Psi^2_1 \Psi^2_2]/4!$, it follows that we also need $v^{}_0>0$ to ensure that both $\Psi_{1,2}$ condense in the ordered phase.

\begin{figure}[tb]
\includegraphics[width=0.75\linewidth]{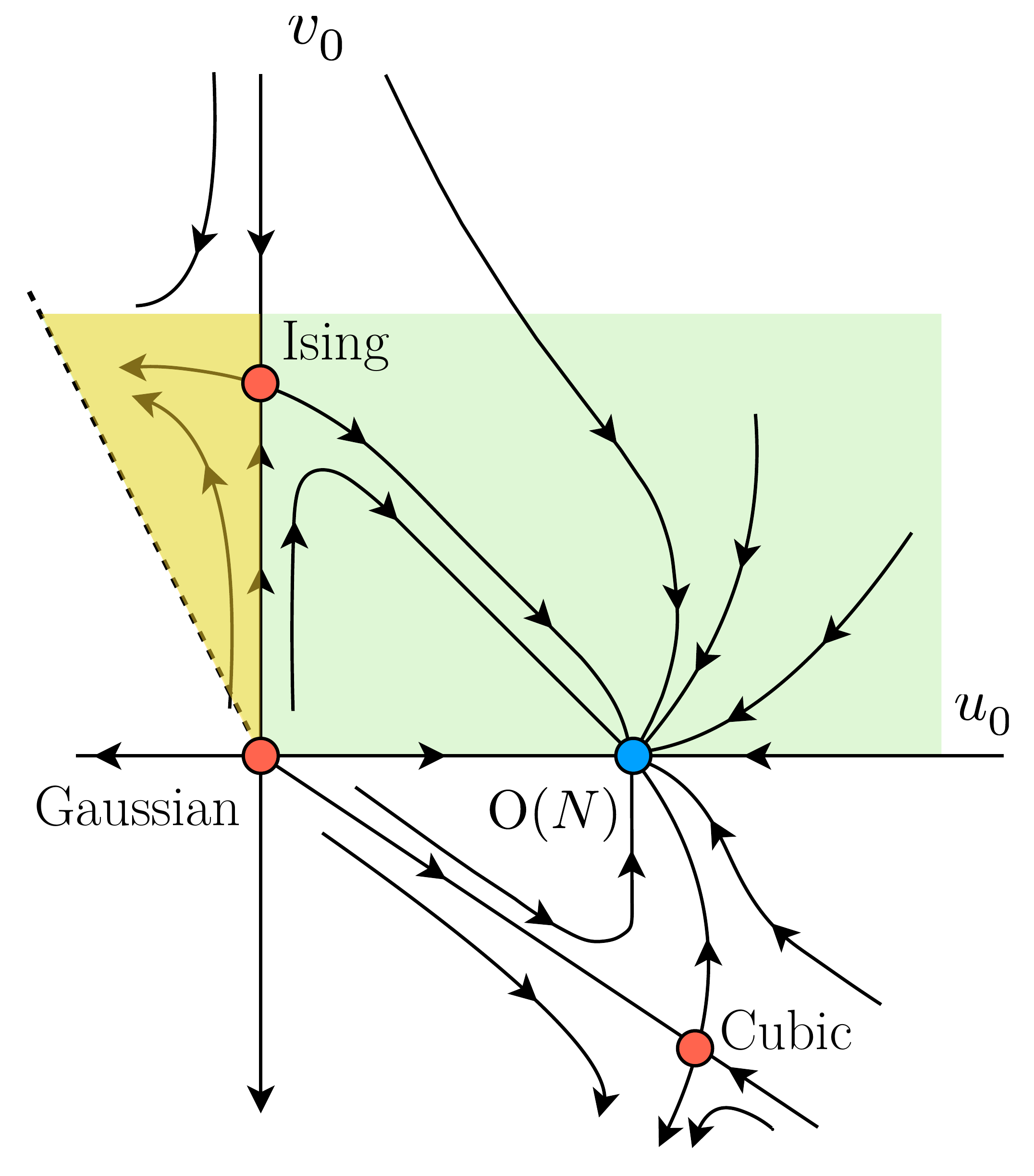}
\caption{RG flow of the theory \eqref{eq:HStr} in the $(u^{}_0, v^{}_0)$ coupling plane for $D$\,$=$\,$3$, $N$\,$=$\,$2$\,$<$\,$N^{}_c$. The blue dot represents the stable $O(N)$-symmetric FPs whereas the other three unstable FPs are marked in red. Owing to the positivity \eqref{eq:stability} and condensation conditions, the parameters of our theory are always constrained to lie in the (green/yellow) region to the right of $v^{}_0$\,$=$\,$-2\, u^{}_0$ (dashed line) in the upper-half plane. The yellow wedge marks the region from which the stable FP is inaccessible. Figure adapted from Ref.~\onlinecite{Pelissetto2000}. }
\label{fig:Flow1}
\end{figure}

The theory defined by the Hamiltonian~\eqref{eq:HStr} has been extensively studied over the past few decades and is known to have four FPs \cite{PhysRevB.8.4270, aharony1976phase}:
\begin{itemize}
\item The trivial Gaussian one at $u^{}_0 = v^{}_0 = 0$.
\item The Ising FP with $u^{}_0, v^{}_0 \ne 0$. At this FP, the Hamiltonian~\eqref{eq:HStr} corresponds to that of $N$ Ising systems coupled by the O($N$)-symmetric interaction. 
\item The $O(N)$-symmetric FP with $u^{}_0 \ne 0 , v^{}_0 = 0$.
\item The cubic FP with $u^{}_0 \ne 0 , v^{}_0 \ne 0$.
\end{itemize}
Both the Gaussian and Ising FPs are unstable for any number of components $N$. For sufficiently small $N$\,$<$\,$N_c$, the O($N$)-symmetric FP is stable while the cubic one is unstable (this designation is reversed for $N$\,$>$\,$N_c$)~\cite{Pelissetto2000}. Importantly, for $D=3$, $N_c$ has been shown to be less than 3 using perturbative expansions~\cite{PhysRevB.56.14428,PhysRevB.61.15136,adzhemyan2019six} and numerical conformal bootstrap~\cite{chester2020bootstrapping}. Hence, in our case, with $D=3$, $N=2$, the O($N$)-symmetric FP describes the generic critical behavior of the system and the resultant RG flow is sketched in Fig.~\ref{fig:Flow1}. 

In the upper-half plane, the only FP is the Ising one which is unstable. 
If $v^{}_0$\,$>$\,$0$ (as required for the simultaneous condensation of $\Psi^{}_{1,2}$) and $u^{}_0$\,$>$\,$0$, at long distances, the theory would therefore flow to the O($N$)-symmetric FP. The transition would be second-order, governed by this stable FP, and one would expect the emergence of an O(2) symmetry at the critical point~\cite{PhysRevLett.99.207203}. The cubic term is a ``dangerously'' irrelevant operator and generates correction to scaling in the ordered phase~\cite{Pelissetto2000}. However, even though the RG flow has a  stable FP, there is a region in the $(u^{}_0, v^{}_0)$ plane, defined by the wedge $\Omega$\,$=$\,$\{(u^{}_0, v^{}_0)$\,$\vert$\,$-v^{}_0/2$\,$<$\,$u^{}_0$\,$<$\,$0\}$, from which the FP is \textit{inaccessible}~\cite{PhysRevB.23.3943}. If $u_0^{}$ were to the left of the separatrix running from the Gaussian to the Ising FP, the flow could not reach the O($N$) FP as the separatrix marks the boundary of the domain of attraction of the stable FP~\cite{bailin1977phase}. Outside the attraction domain of the FPs, the flow goes away towards more negative values of $u_0^{}$ and/or $v_0^{}$, eventually reaching the region where the quartic interaction no longer satisfies the stability condition: these RG trajectories should be related to first-order phase transitions~\cite{Pelissetto2000}. This result is to be contrasted with the mean-field prediction of a continuous transition in the entire stability wedge defined by Eq.~\eqref{eq:stability}.

\begin{figure*}[t]
\includegraphics[width=\linewidth]{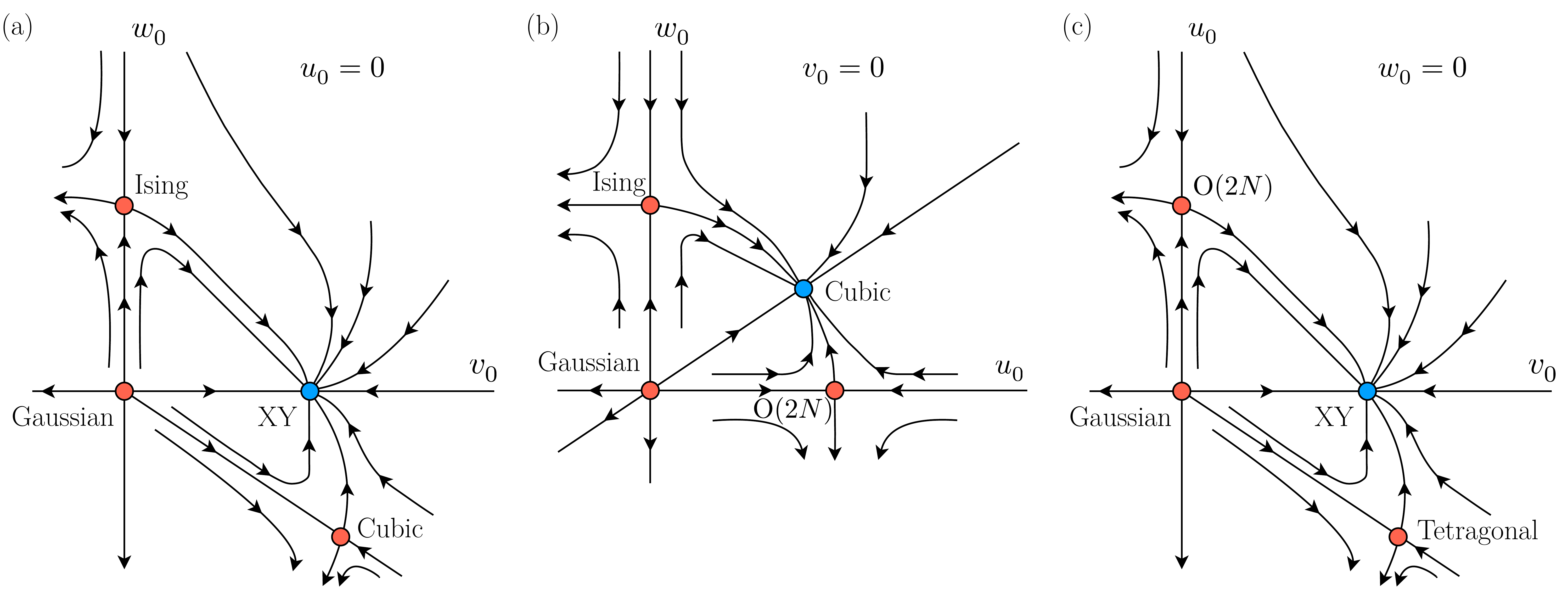}
\caption{Schematic RG flow of the tetragonal theory, described by the LGW Hamiltonian \eqref{Hphi4tetra} with $M$\,$=$\,$2$, $N$\,$>$\,$1$, in the (a) $u^{}_0$\,$=$\,$0$, (b) $v^{}_0$\,$=$\,$0$, and (c) $w^{}_0$\,$=$\,$0$ planes. In the $v^{}_0$\,$=$\,$0$ subspace, the stable FP is the cubic one whereas on the other two planes, the XY FP is stable. As before, all the stable (unstable) FPs are marked in blue (red). Figure adapted from Ref.~\onlinecite{Pelissetto2000}.}
\label{fig:Flow2}
\end{figure*}

\subsubsection{Checkerboard to striated}

The broken symmetry in the striated phase is $\mathbb{Z}_2$\,$\times$\,$\mathbb{Z}_2$\,$\simeq$\,$D_2$ (which is isomorphic to the Klein 4-group). By itself, the checkerboard phase already breaks a $\mathbb{Z}_2$ symmetry. The residual $\mathbb{Z}_2$ symmetry that is further broken on going from the checkerboard to the striated phase is generated by $\{1, T^{}_d\}$, where $T^{}_d$\,$\equiv$\,$T^{}_x T^{}_y$ denotes unit translations along the diagonals ($T_d^2$\,$=$\,$1$). Note that both the checkerboard and striated phases preserve fourfold-rotational symmetry, so the original $D_4$ symmetry of the Hamiltonian \eqref{eq:HStr} is broken down to its subgroup $C_4$.

The striated phase can be distinguished from the checkerboard by defining the order parameter 
\begin{equation}
m \equiv \bigg\lvert \sum_i \left\{ (-1)^{\mathrm{row}(i)} +(-1)^{\mathrm{col}(i)} \right\} \langle n_i \rangle \bigg\rvert,
\end{equation} which measures the differential occupation of sites along the diagonals. Now one can write down a Landau functional in powers and derivatives of the order parameter as usual and the resultant effective Hamiltonian is given by
\begin{equation}
\mathcal{H}_m^{\phantom{\dagger}} = \int d^D \vect{x} \left[ \frac{1}{2}\left\{\left(\partial^{}_\mu m \right)^2 +  r\, m^2 \right\}+ \frac{u^{}_0}{4!}\, m^4 \right],
\end{equation}
which is the same as the $\Phi^4$ field theory studied above in Eq.~\eqref{eq:Phi4}. The $\mathbb{Z}_2$-symmetry-breaking transition from the checkerboard to the striated phase is thus also in the Ising universality class controlled by the Wilson-Fisher FP.

\subsection{The star phase}

To derive the LGW theory for the quantum phase transition from the disordered to the star phase, we begin by noting that the Fourier maxima in the star phase are at $(\pi, 0)$,  $(0, \pi)$, $(\pi/2, \pi)$, and $(\pi, \pi/2)$. However, recognizing that $(\pi, 0)$\,$=$\,$2\, (\pi/2, \pi)$---and similarly for $(0, \pi)$---we can write the magnetization as simply
\begin{alignat}{1}
\rho(\vect{r}) = \mathrm{Re} \left(\Psi_1^{}\, \mathrm{e}^{i \,(\pi/2,\pi)\cdot\vect{r}} + \Psi_2^{}\, \mathrm{e}^{i \,(\pi, \pi/2)\cdot\vect{r}} \right)\,,
\end{alignat}
with $\Psi^{}_1$,\,$\Psi^{}_2$\,$\in$\,$\mathbb{C}$, and the other wavevectors are described by harmonics $\Psi_{1,2}^2$ of the order parameters. Using the basis $\left(\Psi^{}_1,\Psi^{}_2,\Psi^{*}_1,\Psi^{*}_2\right)$, the symmetry operations can be represented by the following matrices:
\begin{alignat}{3}
\nn T^{}_x &= 
\left[
    \renewcommand*{\arraystretch}{1}
    \begin{array}{cccc}
i & 0 & 0 & 0 \\
0 & -1 & 0 & 0 \\
0 & 0 & -i & 0 \\
0 & 0 & 0 & -1 \\
\end{array}\right],\quad
T^{}_y &&= \begin{bmatrix} 
-1 & 0 & 0 & 0 \\
0 & i & 0 & 0 \\
0 & 0 & -1& 0 \\
0 & 0 & 0 & -i \\
\end{bmatrix},
\\
\nn R^{}_x &= 
\begingroup
\setlength\arraycolsep{8pt}
\begin{bmatrix} 
1 & 0 & 0 & 0 \\
0 & 0 & 0 & 1 \\
0 & 0 & 1 & 0 \\
0 & 1 & 0 & 0 \\
\end{bmatrix},\quad
\endgroup
R^{}_y &&= 
\begingroup
\setlength\arraycolsep{7pt}
\begin{bmatrix} 
0 & 0 & 1 & 0 \\
0 & 1 & 0 & 0 \\
1 & 0 & 0 & 0 \\
0 & 0 & 0 & 1 \\
\end{bmatrix},\quad
\endgroup
\\
C^{}_4 &= 
\begingroup
\setlength\arraycolsep{7pt}
\begin{bmatrix} 
0 & 0 & 0 & 1 \\
1 & 0 & 0 & 0 \\
0 & 1 & 0 & 0 \\
0 & 0 & 1 & 0 \\
\end{bmatrix}.
\endgroup
\end{alignat}
These five matrices generate a subgroup of O(4) and the effective Hamiltonian composed of all polynomials invariant under this group  (up to quartic order) is
\begin{align}
\label{eq:L2}
\mathcal{H}^{}_s= \int d^D \vect{x} \Bigg[ &\frac{1}{2} \sum_{i=1}^2\left\{\left\lvert\partial^{}_\mu \Psi^{}_i\right\rvert^2 +  r\, \lvert \Psi_i\rvert^2  \right\}\nn\\
&+ \sum_{i=1}^2 \big[ z^{}_1\, \lvert \Psi^{}_i\rvert^4+  z^{}_3 \left\{\Psi_i^4 + (\Psi_i^*)^4 \right\} \big]\nn\\ 
&+ z^{}_2\, \lvert \Psi^{}_1 \rvert^2  \lvert \Psi^{}_2 \rvert^2 \Bigg].
\end{align}
For stability of the theory, the coefficients $z^{}_i$ must obey the positivity conditions
\begin{equation}
z^{}_1 - 2 \vert z^{}_3 \vert >0 \quad \mbox{and} \quad 2 z^{}_1 + z^{}_2 - 4 \vert z^{}_3 \vert > 0.
\label{eq:stability2}
\end{equation}
We further require $z^{}_2$\,$-$\,$2\,(z^{}_1$\,$-$\,$2\, \lvert z^{}_3\rvert)$\,$>$\,$0$ to ensure that only one of $\Psi_{1,2}$ condenses at a time, as observed in the star phase.

\begin{figure}[tb]
\includegraphics[width=\linewidth]{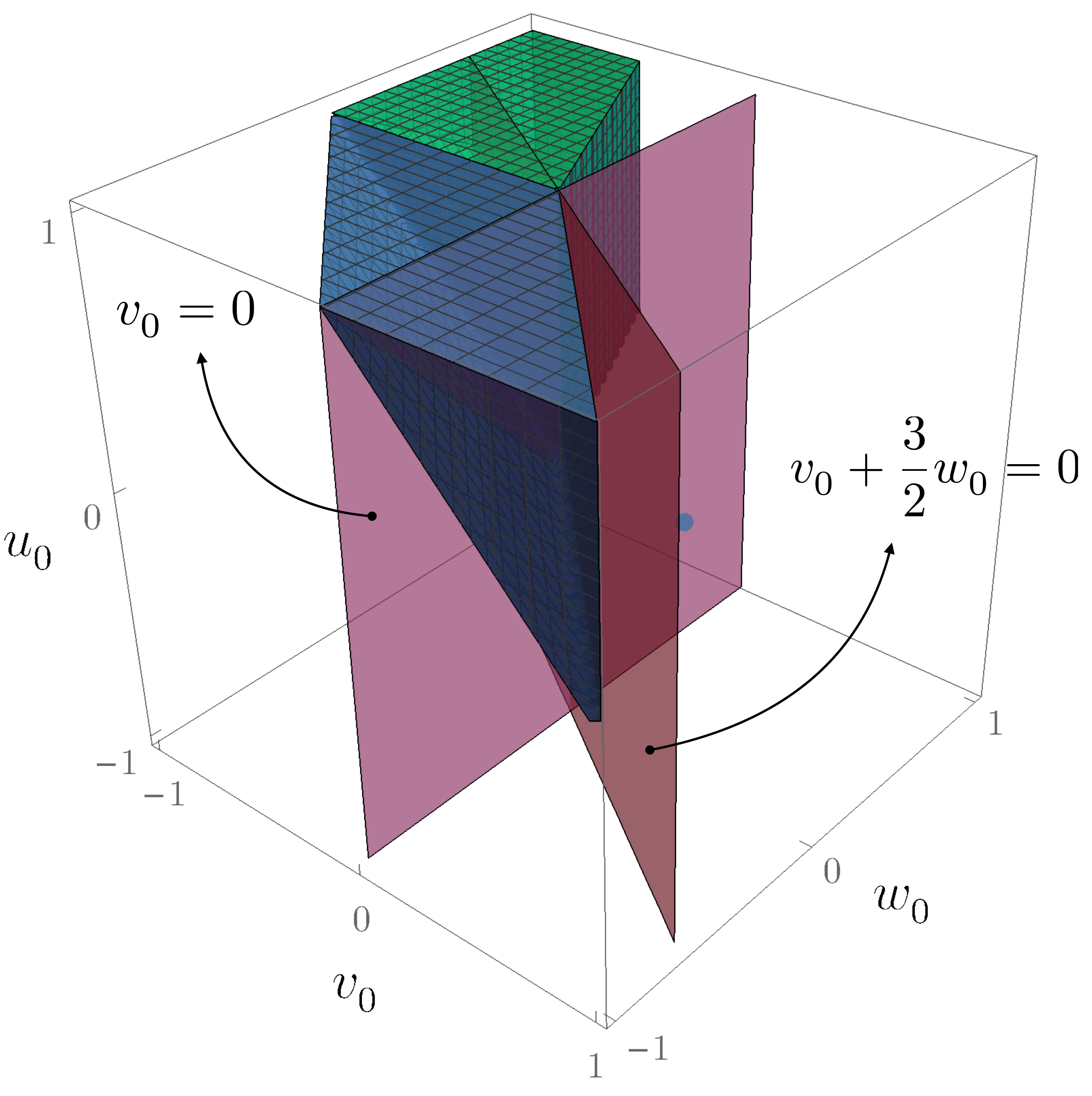}
\caption{\label{fig:region}Allowed region in parameter space of the tetragonal theory \eqref{Hphi4tetra}, as defined by Eq.~\eqref{eq:region} for $w^{}_0 $\,$>$\,$0$ (green) and $w^{}_0$\,$<$\,$0$ (blue). The two planes $v^{}_0$\,$ =$\,$0$ and $v^{}_0$\,$+$\,$(3/2)w^{}_0$\,$ =$\,$ 0$ (on which the cubic FPs are stable) are shaded in pink. The blue dot represents the globally stable XY FP; its $v_0^{}$ coordinate is greatly exaggerated here for the sake of visual clarity.}
\end{figure}

This model is equivalent to the so-called \textit{tetragonal} theory which is the $M$\,$=$\,$2$ version of the general three-coupling LGW Hamiltonian 
\begin{align}
\label{Hphi4tetra}
{\cal H}_\phi^{} = &\int d^D \vect{x} 
\Bigg[ \sum_{i,a}\frac{1}{2}
\left\{ (\partial^{}_\mu \phi^{}_{a,i})^2 + r \phi_{a,i}^2 \right\}\nn\\
&+  \sum_{ij,ab} {\frac{1}{4!}}\left( u^{}_0 + v^{}_0 \delta^{}_{ij} + w^{}_0 \delta_{ij}\delta^{}_{ab} \right)
\phi^2_{a,i} \phi^2_{b,j} \Bigg],
\end{align}
where $a,b$\,$=$\,$1,...M$ and $i,j$\,$=$\,$1,...N$; for our purposes, $N$\,$=$\,$2$. For these parameters ($M,N$\,$=$\,$2$) it corresponds to Eq.~\eqref{eq:Hstar} in the main text. The theory~\eqref{Hphi4tetra} is, of course, constrained to lie within the region of parameter space where Eq.~\eqref{eq:stability2} as well as the abovementioned condition for the mutually exclusive condensation of $\Psi_i^{}$ are both satisfied.
The mapping between the two sets of coefficients is given by $u^{}_0 $\,$=$\,$12\, z^{}_2$, $v^{}_0$\,$=$\,$12\, (2 z^{}_1 - z^{}_2- 12 z^{}_3)$, $w^{}_0 $\,$=$\,$ 192\, z^{}_3$, which implies that the allowed region in $(u^{}_0,v^{}_0,w^{}_0)$ space is defined by Eq.~\eqref{eq:region} in the main text (see Fig.~\ref{fig:region}).

Focusing on $M$\,$=$\,$2$, while keeping $N$\,($>$\,$1$) general, it is instructive to consider certain limits of the Hamiltonian \eqref{Hphi4tetra}.
For $u^{}_0$\,$=$\,$0$, the model reduces to $N$ decoupled cubic models with two-component spins, while for $v^{}_0$\,$=$\,$0$, it describes a cubic model with 2$N$-component spins. In the case $w^{}_0$\,$=$\,$0$, the tetragonal Hamiltonian is equivalent to $N$ coupled XY models \cite{aharony1976phase, PhysRevB.10.892} (also known as the ``$MN$ model'' \cite{grinstein1976application}). These limits, and their combinations, are useful in understanding the different FPs of the theory.
Within the framework of the $\varepsilon$ expansion, the tetragonal model has eight FPs \cite{mukamel1975physical,mukamel1975epsilon,mukamel1976physical}, which are labeled as follows:
\begin{itemize}
\item Gaussian: $u^{}_0 = v^{}_0 = w^{}_0 = 0 $,
\item Ising: $u^{}_0 = v^{}_0 = 0 $; $w^{}_0 \ne 0 $,
\item Heisenberg [$O(2N)$-symmetric]: $u^{}_0 \ne 0$; $v^{}_0 = w^{}_0 = 0 $,
\item XY [$O(2)$-symmetric]: $u^{}_0 = w^{}_0 = 0$; $v^{}_0 \ne 0$,
\item Tetragonal: $u^{}_0 \ne 0$; $v^{}_0 \ne 0$; $w^{}_0 = 0 $, and
\item Cubic: $u^{}_0 \ne 0$; $v^{}_0 = 0 $; $w^{}_0 \ne 0$.
\end{itemize}
One thus obtains six FPs. Additionally, the tetragonal Hamiltonian is symmetric under the transformation \cite{korzhenevskil1976connection}
\begin{align}
&\left (\phi^{}_{1,i}, \phi^{}_{2,i} \right) \rightarrow \frac{1}{\sqrt{2}}\left(\phi^{}_{1,i}+ \phi^{}_{2,i},\phi^{}_{1,i}- \phi^{}_{2,i} \right),\nn\\
&\left(u^{}_0,v^{}_0,w^{}_0 \right) \rightarrow \left(u^{}_0,v^{}_0+\frac{3}{2}w^{}_0, -w^{}_0 \right).
\label{eq:transform}
\end{align}
The two remaining fixed points are obtained by applying the transformation \eqref{eq:transform} to the Ising and cubic FPs listed above, bringing the total to eight. The RG flow of the theory along three different planes (corresponding to one of the quartic couplings being zero) is shown in Fig.~\ref{fig:Flow2}. As described in the main text, the generic critical behavior of the tetragonal theory is described by the XY FP irrespective of $N$.

Crucially, even though the XY FP is globally stable, it is shielded from our allowed region \eqref{eq:region} by the $v^{}_0$\,$ =$\,$0$ and $v^{}_0$\,$+$\,$(3/2)w^{}_0$\,$ =$\,$ 0$ planes (see Fig.~\ref{fig:region}).
We can now extend the arguments of \citet{PhysRevB.23.3943} outlined above to our three-dimensional parameter space. Suppose our initial conditions place the microscopic theory in a regime where $w^{}_0$\,$>$\,$-(2/3)v^{}_0$\,$>$\,$0$ (rightmost green region of Fig.~\ref{fig:region}). All points in this region are separated from the XY FP by the $v^{}_0 = 0$ plane. On this plane, we know that the cubic FP is stable but its stability matrix
possesses a negative eigenvalue in full parameter
space \cite{PhysRevE.64.047104}. Hence, given these initial conditions, the RG flow would take one \textit{away} from the $v^{}_0$\,$ =$\,$0$ plane, and accordingly, the stable XY FP. Using the analogous properties of the $v^{}_0$\,$+$\,$(3/2)w^{}_0$\,$ =$\,$ 0$ plane, we can generalize this argument to \textit{all} possible initial conditions shown in Fig.~\ref{fig:region}. As a result, we can conclude that the stable XY FP is inaccessible starting from our allowed region of parameter space, rendering the transitions from the disordered to the star phase first-order.

Lastly, let us mention that besides the eight FPs  referenced above, the tetragonal theory~\eqref{Hphi4tetra}---restricted to the $w^{}_0$\,$=$\,$0$ plane---is believed to have another stable FP in the region
$v_0^{}$\,$<$\,$0$, $u_0^{}$\,$>$\,$0$, the presence of which is not directly predicted by the $\varepsilon$-expansion framework: this is the O($2$)$\times$O($2$)-symmetric \textit{chiral} FP~\cite{pelissetto2005interacting}. While $\varepsilon$-expansion~\cite{kawamura1998universality} and fixed-dimension~\cite{PhysRevB.63.140414} calculations disagree on the existence of a stable FP corresponding to this chiral universality class for $N$\,$=$\,$2$ in $D$\,$=$\,$3$, we note that within the RG approach, fluctuation-induced first-order transitions are always still possible for systems lying outside the attraction domain of the chiral FP if it exists~\cite{PhysRevB.66.180403}. Moreover, it is presently unclear whether this chiral FP is stable in the enlarged tetragonal theory with $w^{}_0$\,$\ne$\,$0$.
	
\bibliography{bibliography.bib}
	
\end{document}